\pdfoutput=1
\documentclass[a4paper,fleqn,usenatbib]{aa}

\usepackage{graphicx}
\usepackage{txfonts}
\usepackage{natbib}
\usepackage{multirow}
\usepackage[]{hyperref}

\hypersetup{
    bookmarks=true,                                   
    pdftitle={}, 
    pdfauthor={}, 
    colorlinks=true,                                  
    linkcolor=blue,                                   
    citecolor=blue,                                   
    urlcolor=blue                                     
}

\begin{document}

\title{First evidence of inertial modes in $\gamma$ Doradus stars} 
\subtitle{The core rotation revealed}

\author{R-M. Ouazzani\inst{1}, F. Lignières\inst{2}, M-A. Dupret\inst{3}, S.J.A.J. Salmon\inst{3}, J. Ballot\inst{2}, S. Christophe\inst{1}, and M. Takata\inst{4}}

\institute{LESIA, Observatoire de Paris, Université PSL, CNRS, Sorbonne Université, Université de Paris, 5 place Jules Janssen, 92195 Meudon, France
  \and
  IRAP, Université de Toulouse, CNRS, UPS, CNES, 14 Avenue Édouard Belin, 31400, Toulouse, France
  \and
 STAR Institute, Université de Liège, Allée du 6 Août 19, 4000 Liège, Belgium 
  \and
Department of Astronomy, School of Science, The University of Tokyo, 7–3–1 Hongo, Bunkyo-ku, Tokyo 113–0033, Japan
}
\date{Draft: \today; Received xxx; accepted xxx}

\abstract{The advent of space photometry with CoRoT and \textit{Kepler} has allowed the gathering of exquisite and long time series for a wealth of main sequence stars, including $\gamma$ Doradus stars whose detailed seismology was unachievable from the ground.}
         {$\gamma$ Doradus stars present an incredibly rich pulsation spectra, with gravito-inertial modes, in some cases supplemented with $\delta$ Scuti-like pressure modes -for the \textit{hybrid} stars- and in numerous cases with Rossby modes. The present paper aims at showing that, in addition to these modes established in the radiative envelope, pure inertial modes, trapped in the convective core, can be detected in \textit{Kepler} observations of $\gamma$ Doradus stars, thanks to their resonance with the gravito-inertial modes.}
         {We start by using a simplified model of perturbations in a full sphere of uniform density. Under these conditions, the spectrum of pure inertial modes is known from analytical solutions of the so-called Poincaré equation. We then compute coupling factors which help select the pure inertial modes which interact best with the surrounding dipolar gravito-inertial modes. Using complete calculations of gravito-inertial modes in realistic models of $\gamma$ Doradus stars, we are able to show that the pure inertial/gravito-inertial resonances appear as \textit{dips} in the gravito-inertial mode period spacing series at spin parameters close to those predicted by the simple model. We find the first evidence of such dips in the \textit{Kepler} $\gamma$ Doradus star KIC5608334. Finally, using complete calculations in isolated convective cores, we find that the spin parameters of the pure inertial/gravito-inertial resonances are also sensitive to the density stratification of the convective core.}
         {In conclusion, we have discovered that certain \textit{dips} in gravito-inertial mode period spacings observed in some \textit{Kepler} stars are in fact the signatures of resonances with pure-inertial modes that are trapped in the convective core. }
         {This holds the promise to finally access the central conditions , i.e. rotation and density stratification, of intermediate-mass stars on the main sequence.}


\titlerunning{The core rotation revealed by inertial modes in $\gamma$ Doradus stars}
\authorrunning{Ouazzani et al.}

\maketitle

\section{Introduction}
Angular momentum transport in stellar radiative zones, and its evolution with time is a major open issue in stellar physics. Observations of stars in evolved stages provided by the {\it Kepler} mission \citep{Beck2012,Deheuvels2012,Mosser2012a} have proved our then current models of transport wrong \citep{Eggenberger2012,Marques2013}, and the various attempts to invoke additional mechanisms \citep{Cantiello2014,Fuller2014,Belkacem2015b,Pincon2017,Mathis2018,Fuller2019} have not solved this issue, or are still matter of debate.

With the purpose of bringing a main-sequence counterpart to the seismic constraint of rotation in evolved stars, $\gamma$ Doradus stars g-modes monitored by {\it Kepler} have revealed near-core rotation rates in stars which are progenitors of red giants \citep{VanReeth2016,Christophe2018,Li2019}. This has allowed the test of angular momentum transport models against near-core rotation measurements in main-sequence intermediate-mass stars. As a result, \cite{Ouazzani2019} have found that the rotational transport as formalized by \cite{Zahn1992} does not allow them to reproduce observations of $\gamma$ Doradus stars. Hence, as for the more evolved stages, another -yet to be found- mechanism seems to transport angular momentum from the innermost part of the radiative zone outwards. Whether it leads to rigidification of the rotation profile still remains unproved, but measurements in a handful of stars seem to suggest so \citep{Kurtz2014,Saio2015,Murphy2016,VanReeth2018,Li2019}.

The constraint of the full internal rotation profile in intermediate-mass stars would be extremely valuable, as it would answer this question and allow further characterization of the transport mechanism of angular momentum at stake. Once again, $\gamma$ Doradus stars could help bring such constraint. In addition to the g-modes period spacing series, which interpretation allowed near-core rotation measurements, these stars harbor pulsations in Rossby modes \citep{Saio2018a,Li2019} which hopefully bring rotational information in a different cavity than g-modes (\citealt{VanReeth2018}, Christophe et al. in prep). The present paper demonstrates the existence of a signature of pure inertial modes trapped in the convective core within g-modes period spacing series, holding the possibility to probe the rotation rate of the convective core. 

Inertial waves are caused by the Coriolis acceleration, their frequency in the rotating frame ranging from $0$ to $2\Omega$. One usually refers to pure inertial waves when other restoring such as buoyancy or pressure forces are absent or negligible. This is relevant for slow motions in stellar convective zones, where turbulent heat transport maintains an isentropic mean stratification or equivalently a vanishing Brunt-V\"ais\"al\"a frequency $N$. We thus expect convective zones and in particular the convective cores of massive and intermediate-mass stars to harbor pure inertial waves. Pure inertial modes in a full sphere have been studied for a fluid of constant density \citep{Bryan1889, Greenspan1968,Rieutord1991} and for isentropic polytropes \citep{Papaloizou1981, Lee1992, Dintrans2001, Lockitch1999, Wu2005}.  Pure inertial modes trapped in the solar convective zone have been investigated by \citet{Guenther1985} and \cite{Dziembowski1987b}. The coupling of pure inertial modes in stellar convective zones with gravito-inertial modes in adjacent radiative zones has been studied in the case of tidally excited modes in solar-type stars \citep{Ogilvie2004} and in massive stars  \citep{Papaloizou1997}. \citet{Lee1987a} computed oscillatory convective modes present in a rotating convective core with a slightly superadiabatic stratification coupled with gravito-inertial modes or Rossby modes in the radiative enveloppe. 

The present article holds the first unambiguous theoretical demonstration of the existence of resonances between gravito-inertial modes in the radiative zone, and pure inertial modes in the convective core in $\gamma$ Doradus stars. By comparing with a $\gamma$ Doradus star observed by the {\it Kepler} mission, we show the first observational signature of pure inertial modes. We start by introducing inertial modes in the case of a sphere of uniform density (Sect. \ref{Ss:equations}) and then propose a simple model to describe their coupling with gravito-inertial modes (Sect. \ref{Ss:coupling}). In Sect.\ref{S:synthspectra}, we then explore the occurrence of such gravito-inertial$/$pure inertial resonances in complete calculations. Finally we investigate the dependency of these resonances on stellar characteristics (Sect.\ref{S:stratif}), before investigating the diagnosis potential of these resonances regarding the measurement of rotation in the convective core.

\section{Theoretical spectrum of pure inertial modes}
\label{S:theory}

Although inertial modes in a spherical shell with solid boundaries present complex characteristics such as singular modes in the inviscid limit \citep{Rieutord2018}, inertial modes in the full sphere are expected to be smooth and, in the case of a uniform density sphere, are amenable to analytical solutions \citep[e.g.][]{Greenspan1968}. 

Following \cite{Wu2005}, it is straightforward to show that, in an isentropic medium (where the Brunt-Vaissala frequency $N=0$), small-amplitude adiabatic perturbations show a time and azimuthal (angle $\phi$) dependence $\propto \exp(im\phi + i \omega t)$, and are governed by the wave equation :
\begin{equation}
\label{eq:wave}
\nabla^2 \Psi - s^2 \frac{\partial^2 \Psi }{\partial z^2} = \frac{1}{H}\left(\frac{\partial \Psi}{\partial r} - s^2 \cos \theta \frac{\partial \Psi}{\partial z} +\frac{m s}{r} \Psi \right) - (1-s^2) \frac{\omega_{co}^2}{c_s^2} \Psi
\end{equation}
\noindent where $s=(2 \Omega)/\omega_{co}$ is the spin parameter, that is the ratio of twice the rotation rate $\Omega$ (which is taken uniform here) to the frequency in the co-rotating frame $\omega_{co}$. $\Psi$ is related to the Eulerian pressure perturbation $P'$ and the equilibrium density $\rho_0$ by $\Psi = P'/(\varrho_0 \omega_{co}^2)$, $H=-\varrho_0/(d \varrho_0/d r)$ is the density scale height, and $c_s=\sqrt{\Gamma_1 P_0/\rho_0}$ is the sound speed, here defined by the help of the adiabatic exponent $\Gamma_1$ and the equilibrium pressure $P_0$. Perturbations of the gravitational potential have been neglected through the Cowling approximation.
Away from the star surface, and as long as the inertial frequencies are much lower than pressure wave frequencies, the last term can also be neglected.

In the next two subsections, we first present properties of inertial modes relevant for our purpose and then construct a simple model for the occurrence of gravito-inertial/inertial mode resonance. 

\begin{figure*}[t!]
  \centering
  \includegraphics[width=0.25\linewidth]{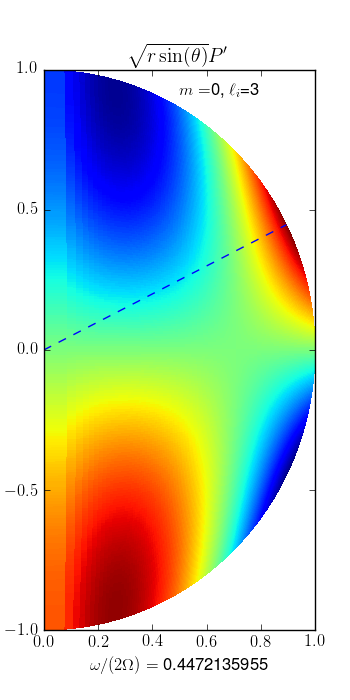}\hspace*{0.5cm}
  \includegraphics[width=0.25\linewidth]{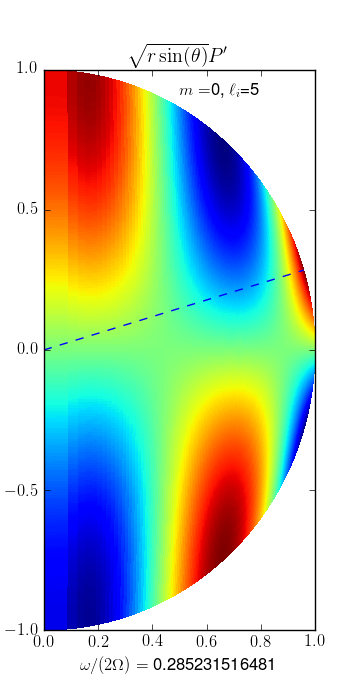}\hspace*{0.5cm}
  \includegraphics[width=0.25\linewidth]{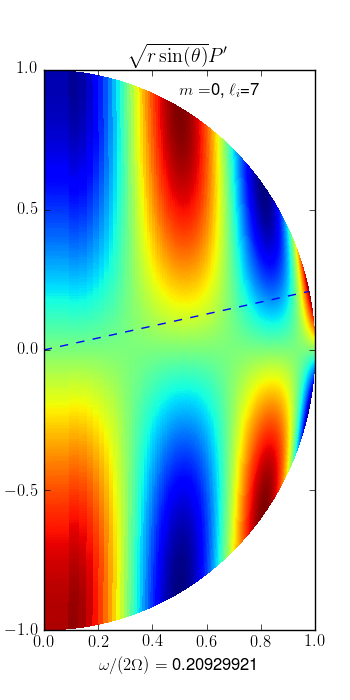}
        \caption{\label{fig:GI_HomoSPhere} Inertial modes in a uniform density sphere. Each of these modes have a strong spatial matching with the dipolar gravito-inertial mode of the same spin parameter. The dotted line specifies the critical latitude $\theta_c = \arccos(\omega/(2\Omega))$.}
\end{figure*}

\subsection{Inertial modes in a full sphere of uniform density}
\label{Ss:equations}

In the absence of density stratification, $H=+\infty$, the wave equation reduces to the so-called Poincaré equation $ \nabla^2 \Psi - s^2 \frac{\partial^2 \Psi }{\partial z^2} =0$. 
This equation is separable using the ellipsoidal coordinates $(x_1,x_2,\phi)$ defined in Appendix~\ref{app:coord}, and the solution, already obtained by \cite{Bryan1889}, reads
\begin{equation}
\Psi(\vec x) = P_{\ell_i}^{m}(x_1) P_{\ell_i}^{m}(x_2)\exp(im\phi + i \omega t)
\end{equation}
where $P_{\ell_i}^{m}$ is the associated Legendre polynomial of degree $\ell_i$. For a given pair $(\ell_i,m)$, the solution $\Psi(\vec x)$ also holds a frequency dependency through the endpoints of the meridional coordinate intervals, $x_1 \in [\mu, 1]$ and $x_2 \in [-\mu, \mu]$, where $\mu=\omega_{co}/(2 \Omega)=1/s$.
In addition to regularity conditions at the center and along the rotation axis, the boundary condition $\xi_r=0$, applied at the sphere surface, determines the inertial mode frequencies as the non-trivial positive roots
of :
\begin{equation}
\label{freq}
\frac{d P_{\ell_i}^{m}}{d \mu} (\mu) = \frac{m}{1 -\mu^2} P_{\ell_i}^{m}(\mu).
\end{equation}

\noindent The detail of the calculations together with a useful discussion on the inertial mode properties can be found in \cite{Wu2005}.

Here we focus on the inertial modes which may interact resonantly with dipolar gravito-inertial modes,
that is, for symmetry reasons, anti-symmetric zonal ($m=0$) modes, symmetric prograde ($m=-1$) and retrograde ($m=1$) modes. For each symmetry class, the set of inertial eigenfrequencies is dense in the $[0, 2 \Omega]$ interval. However, we shall mostly select low degree $\ell_i$ modes. 
For each pair $(\ell_i,m)$, there are $[\ell_i-|m| - \epsilon(\ell_i-|m|)]/2$ eigenfrequencies, where $\epsilon(\ell_i-|m|)=0$ for even modes and $\epsilon(\ell_i-|m|)=1$ for odd modes. For example, for anti-symmetric zonal ($m=0$) modes, one mode is associated with $\ell_i =3$, two modes with $\ell_i =5$, three modes with $\ell_i =7$, and so forth. All modes of a given pair $(\ell_i,m)$ have the same number of nodes along a meridian line at the outer sphere, namely $\ell_i-|m|$ nodes. 

In table~\ref{table:1}, the spin parameters $s$ of the first low $\ell_i-|m|$ eigenmodes that can potentially couple with dipolar gravity modes are listed in ascending $\ell_i-|m|$.
The spatial distributions of the three $(\ell_i =3,5,7; m=0)$ modes of the first column in the table are displayed on Fig.\ref{fig:GI_HomoSPhere}. As we shall see below, these modes are the most likely to couple with a dipolar zonal mode above the convective core.
The remaining three axisymmetric modes of table~\ref{table:1}, one $(\ell_i =5; m=0)$ mode and two $(\ell_i =7; m=0)$ modes are shown 
on Fig.\ref{fig:GI_HomoSPhere_App} of Appendix~\ref{app:modes}. As apparent from these figures, and as mentionned before, the different modes associated with a pair $(\ell_i,m)$ have the same number of nodes along the meridian line at the outer surface but they differ by the number of nodes -on that outer surface meridian line- that are located between the equator and the critical colatitude $\theta_c = \arccos (\mu) = \arccos (1/s)$. For example, the three $(\ell_i,m) =(7,0)$ modes have either no node, one node or two nodes between the equator and the critical colatitude $\theta_c$.

 
\subsection{Coupling with gravito-inertial modes}
\label{Ss:coupling}

Small-wavelength WKB analysis tells that gravity waves can propagate in regions where $\omega < N$ as long as their frequency is sub-inertial, i.e. $\omega < 2 \Omega$ \citep[see for instance][]{Unno1989}. Low-frequency gravity waves can thus propagate from a radiative zone down to a convective core \citep{Dintrans2000,Prat2016}.
If conditions for positive interferences are met, such waves can produce mixed gravito-inertial/pure inertial modes.

Because the spectrum of pure inertial modes is dense in $[0, 2 \Omega]$, there is always a possible frequency match between a sub-inertial gravity mode in the radiative zone and a pure inertial mode in the core. However, a resonant coupling is expected if there is matching between both their frequencies and their spatial distributions at the convective core boundary. In the well-known case of mixed gravity/pressure modes in slowly rotating evolved stars, the spatial matching is ensured by considering modes with the same spherical harmonic degree. There is no such simplification here as the latitudinal distribution of both pure inertial modes and gravito-inertial modes in general depends on the frequency.


\begin{table}
\caption{Inertial-mode spin parameters and coupling coefficients $\zeta_I^{GI}$ (in parenthesis) with the dipolar Hough function of the same symmetry class and spin parameter}              
\label{table:1}      
\centering                                      
\begin{tabular}{c l l l}          
\hline\hline                        
$\ell_i$ & $m=0$ & & \\    
\hline                                   
    3 & 2.2361 (0.62)   &  & \\      
    5 & 3.5059  (0.51) & 1.3071  (0.014) &  \\
    7 & 4.7778  (0.44) & 1.6900 (3.3 10$^{-3}$) & 1.1471 (1.6 10$^{-4}$) \\
\hline \hline                                            
$\ell_i$ & $m=-1$ & & \\    
\hline                                   
    3 & 11.3245  (0.50) &  & \\      
    5 & 29.3302 (0.39) & 1.6900 (4.3 10$^{-3}$) &  \\
    7 & 55.3317 (0.34) & 2.3182 (3.8 10$^{-5}$) & 1.2926 (1.89 10$^{-5}$)   \\
\hline \hline                                            
$\ell_i$ & $m=1$ & & \\    
\hline                                   
    3 & 1.3246 (0.71) & & \\      
    5 & 1.9128 (0.60) & 1.1074 (0.055) & \\
    7 & 2.5309 (0.52) & 1.3570 (0.033) & 1.0546 (1.7 10$^{-3}$) \\ 
\hline
\end{tabular}
\end{table}

Although a detailed mathematical model of the resonance conditions is beyond the scope of the present paper, we propose here a simplified approach to estimate the occurrences of strong resonant coupling between a pure inertial mode and a gravito-inertial mode, respectively  oscillating in the convective core and above it in the radiative zone. We first define a coupling coefficient $\zeta_{I}^{GI}$ as :
\begin{equation}
\label{couple}
\zeta_{I}^{GI} = \frac{\int_0^{\pi/2} \sin\theta \; {P'}_I(r_c, \theta) {P'}_{GI}(r_c, \theta)  d\theta}{\left(\int_0^{\pi/2} \sin\theta \; {P'}_I^2 d \theta \right)^{1/2}  \left(\int_0^{\pi/2} \sin\theta \; {P'}_{GI}^2 d \theta \right)^{1/2}}
\end{equation}
\noindent where $r_{c}$ is the radius of the convective core boundary and, ${P'}_I$ and ${P'}_{GI}$ are the latitudinal pressure profile of the inertial and gravito-inertial modes respectively. 

To compute this coefficient we make the following assumptions : we assume that the latitudinal variations of the gravito-inertial mode is given by the Hough function, i.e. the expected form in the framework of the traditional approximation. For the present work, Hough functions are numerically computed by solving Laplace's tidal equation projected on the basis of spherical harmonics truncated to the degree $\ell_{\mathrm{max}}=500$. The linear system to solve is then described in \citet{Unno1989}. The latitudinal form of the pure inertial modes results from two assumptions. The first one is to neglect the effect of the density stratification. The second is to use the condition $\xi_r=0$ at the convective core radius as a boundary condition to quantize the inertial modes. The inertial modes then take the analytical form given in the previous sub-section. In particular, the pressure perturbation profile at the core radius ${P'}_I$ is simply $\propto P_{\ell_i}^{m}(\cos\theta)$.

\begin{figure}[t!]
  \centering
  \includegraphics[width=1\linewidth]{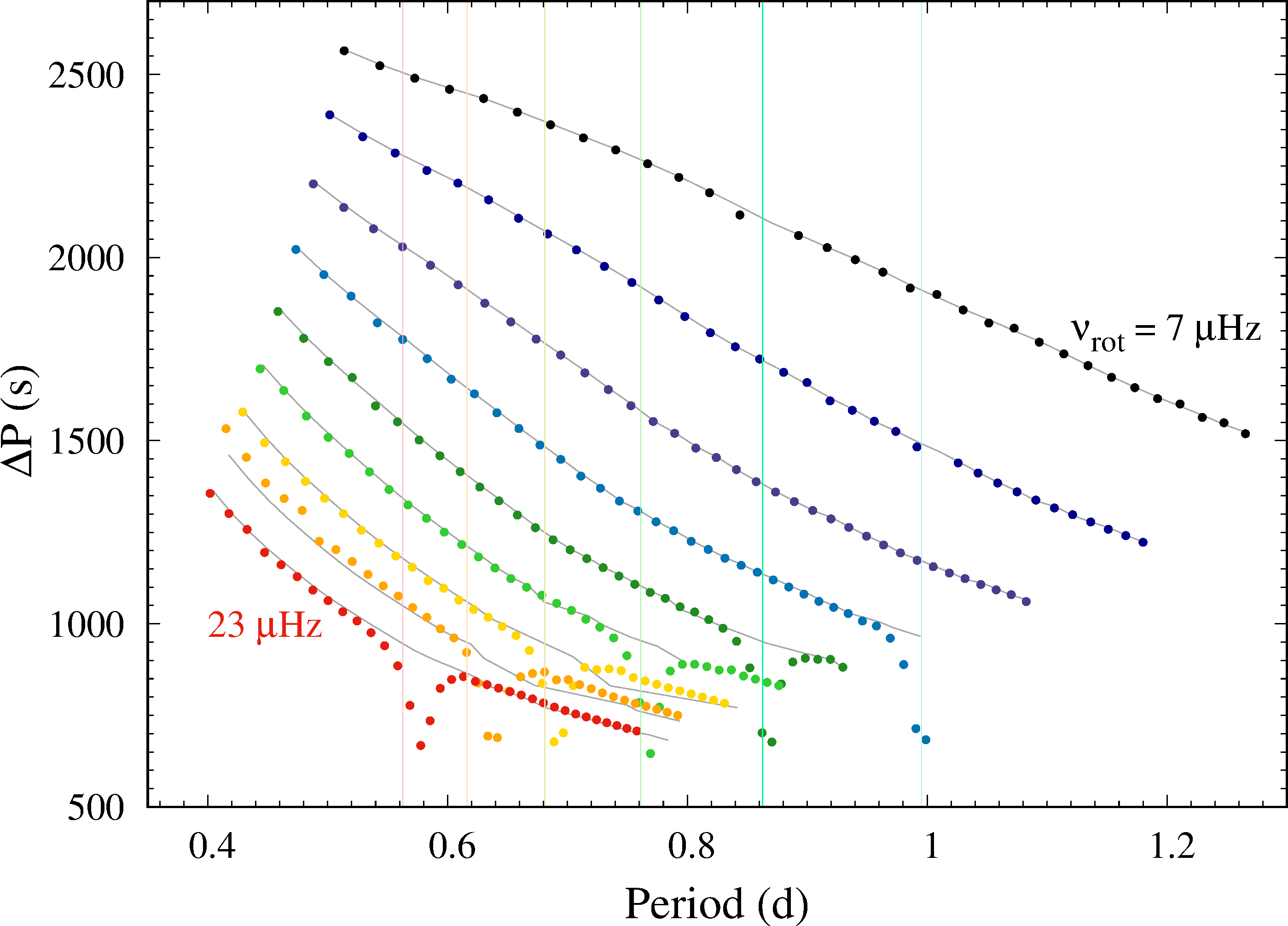}
        \caption{\label{fig:DeltaP_m1.40_allrot} Period spacing as a function of the period for model 1z with different value of the uniform rotation frequency, ranging from 7 $\mu$Hz to 23 $\mu$Hz, by increment of 2 $\mu$Hz, for zonal dipolar modes. Both the period spacings and the periods are shown in the inertial frame of reference. The filled circles are obtained with non-perturbative calculations, while the grey curves are obtained with the traditional approximation of rotation. The vertical lines give the location of the periods of pure inertial modes associated with different rotation rates (only for $\nu_{rot} = 13$ to $23 \mu$Hz), computed for an incompressible inviscid spherical box as described in Sect.\ref{Ss:coupling}.}
\end{figure}

Using Eq.~\ref{couple}, we compute the coupling coefficients $\zeta_{I}^{GI}$ between the uniform-density low-order inertial modes listed in Tab. \ref{table:1} and the dipolar Hough functions corresponding to the same spin parameters. We find that the coupling coefficients are either significant or very small. This property can be readily understood from the geometry of the modes involved. Considering the axisymmetric case, we know that the axisymmetric dipolar Hough functions are single-signed between the equator and the critical colatitude $\theta_c$ corresponding to the spin parameter, and then rapidly goes to zero at higher latitudes \citep{Townsend2003}. Thus multiplying them by an inertial mode which also have no node between the equator and $\theta_c$ and integrating over latitude will give a significant normalized coefficient. By contrast, the coupling with inertial modes that have nodes between the equator and $\theta_c$ will be necessarily much smaller. This geometrical interpretation explains why the inertial modes displayed in Fig.~\ref{fig:GI_HomoSPhere} have significant values of $\zeta_{I}^{GI}$ with the axisymmetric dipolar gravito-inertial mode of equivalent spin parameter and why those displayed in Fig.~\ref{fig:GI_HomoSPhere_App} have very low $\zeta_{I}^{GI}$. Similar geometrical interpretations hold for the coupling coefficient of the prograde and retrograde modes.

Thus, considering dipolar gravito-inertial modes scanning a given frequency range, this simplified model predicts strong coupling with a pure inertial mode at the spin parameters listed in Tab.~\ref{table:1}. The assumptions of the present model will be discussed later in the paper.


\section{Occurrence of gravito-inertial resonances in complete calculations}
\label{S:synthspectra}

\begin{figure}[t!]
  \centering
  \includegraphics[width=1\linewidth]{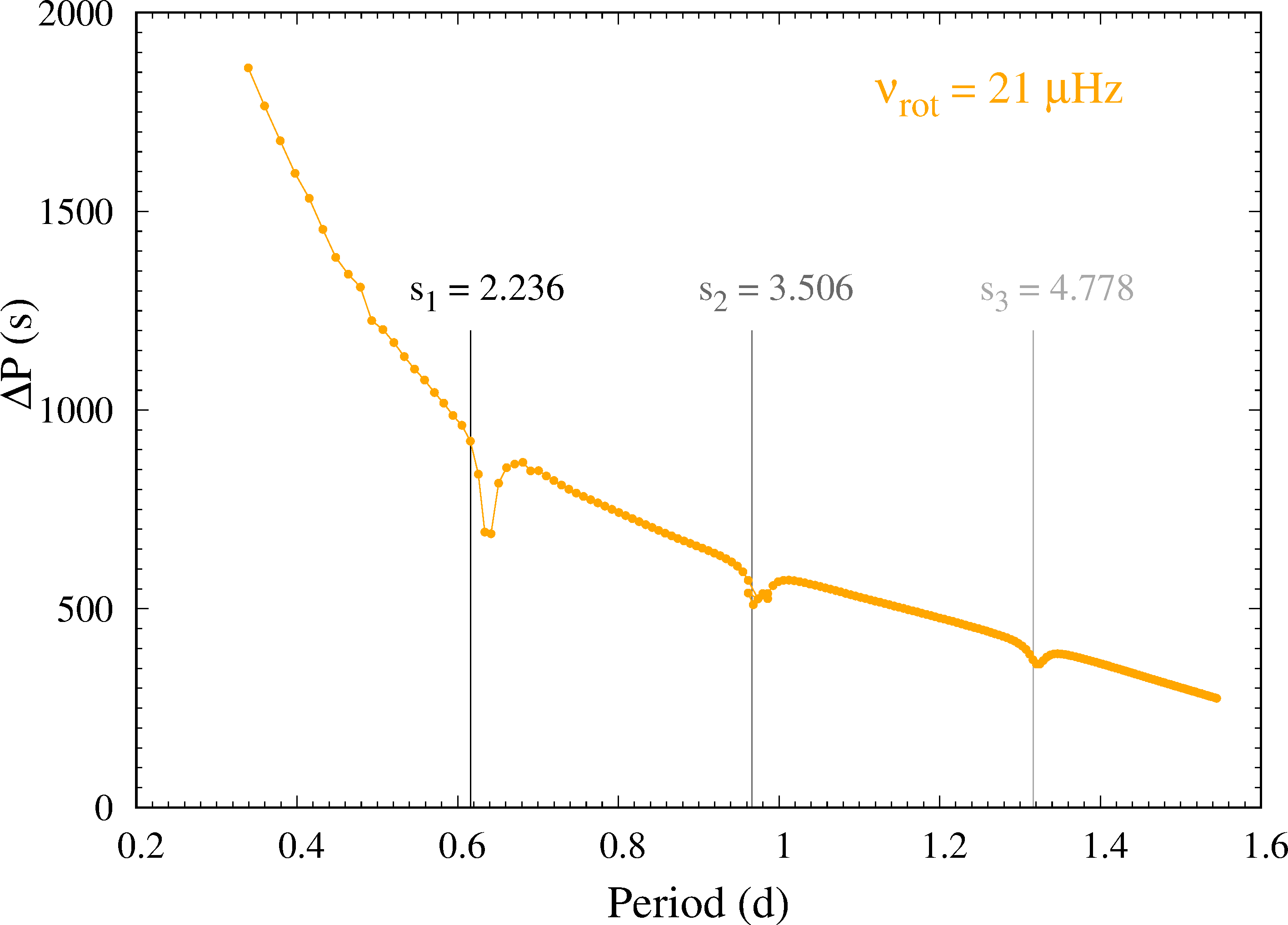}
        \caption{\label{fig:DeltaP_m1.40_21muHz} Period spacing as a function of the period for model 1z for a uniform rotation frequency of 21 $\mu$Hz. Both the period spacings and the periods are shown in the inertial frame of reference.}
 \end{figure}

In this section, we compute oscillation spectra of realistic models of main-sequence intermediate-mass stars. We wish to investigate further the behavior of the mixed gravito-inertial/pure inertial modes, while avoiding the assumptions made in the previous section.

\subsection{Stellar and pulsation models}
\label{Ss:models}
Stellar models were computed with the stellar evolution code {\sc cles} \citep{Scuflaire2008b} for masses between 1.4 and 1.86 M$_{\odot}$, and with initial helium mass fraction $Y = 0.27$. We adopted the AGSS09 metal mixture \citep{Asplund2009} and corresponding opacity tables obtained with  OPAL opacities  \citep{Iglesias1996},  completed  at  low  temperature ($log T <4.1$) with \cite{Ferguson2005} opacity tables. We  used  the  OPAL2001  equation  of  state  \citep{Rogers2002}  and  the  nuclear  reaction  rates  from  NACRE  compilation \citep{Angulo1999}, except  for  the $^{14}N(p,\gamma) ^{15}O$  nuclear  reaction, for which we adopted the cross-section from \cite{Formicola2004}. Surface boundary conditions at $ T = T_{\rm eff}$ were provided by ATLAS 9 model atmospheres \citep{Castelli2003}. Convection was treated using the mixing-length theory (MLT) formalism \citep{Bohm-Vitense1958}  with a parameter $\alpha_{\textrm{MLT}} = 1.70$, close to a solar calibration.

We considered models with turbulent diffusion. Since the {\sc cles} code does not include effects of rotation on transport of angular momentum or chemical species, we instead introduced mixing by turbulent diffusion, following the approach of \citet{Miglio2008}. This reproduces an effect of rotationally-induced mixing that is quite similar to overshooting, but in addition tends to smooth chemical composition gradients inside the star. The coefficient of turbulent diffusion was set to $D_{\textrm{t}}=700 cm^2/s$ and kept constant to this value during evolution and in every layer of the models. This value was selected from a previous calibration to Geneva models with similar masses \citep{Miglio2008}.
\begin{figure*}
 \includegraphics[width=0.31\linewidth,angle=0]{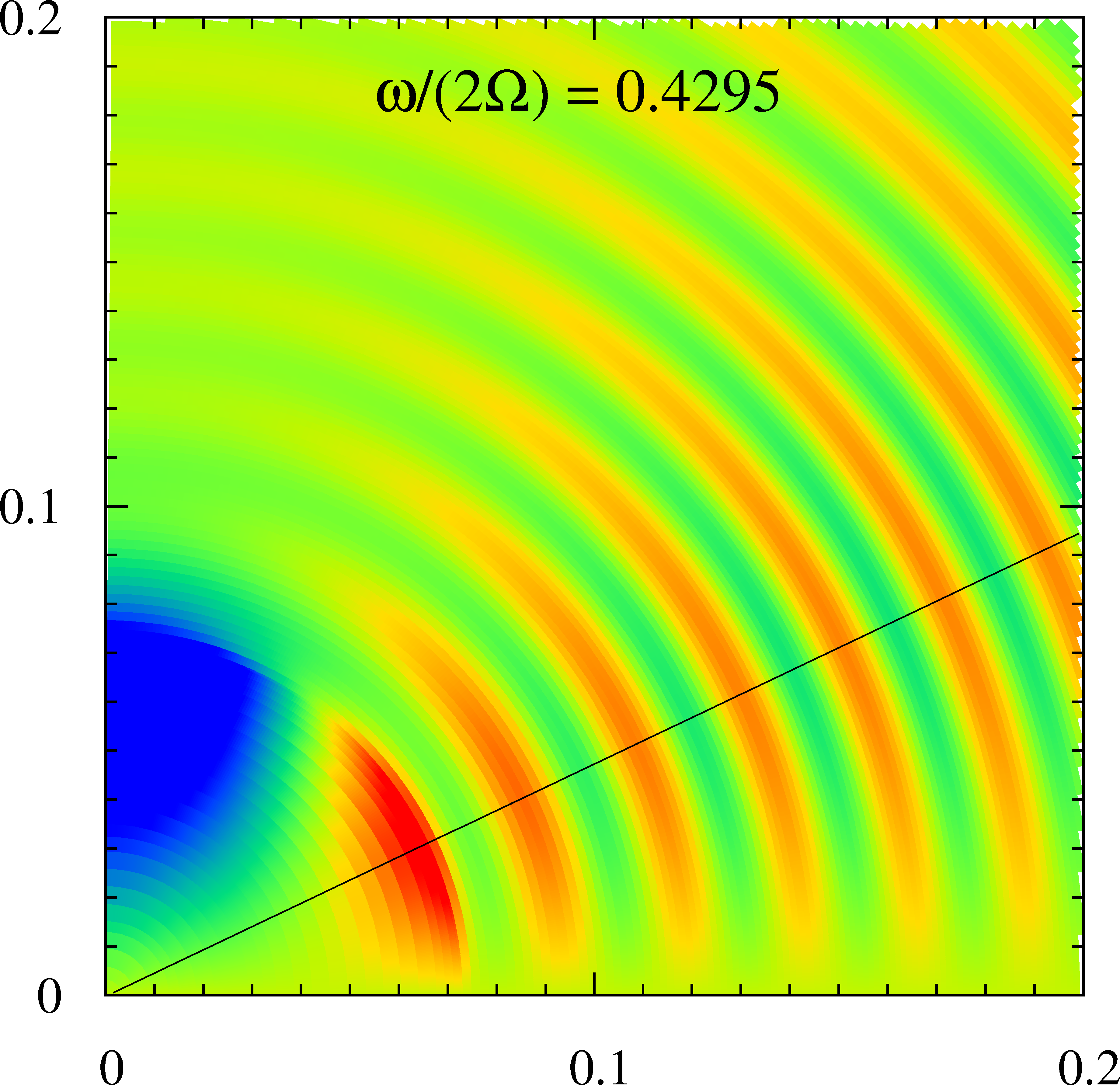} \hspace*{0.cm}  \includegraphics[width=0.31\linewidth]{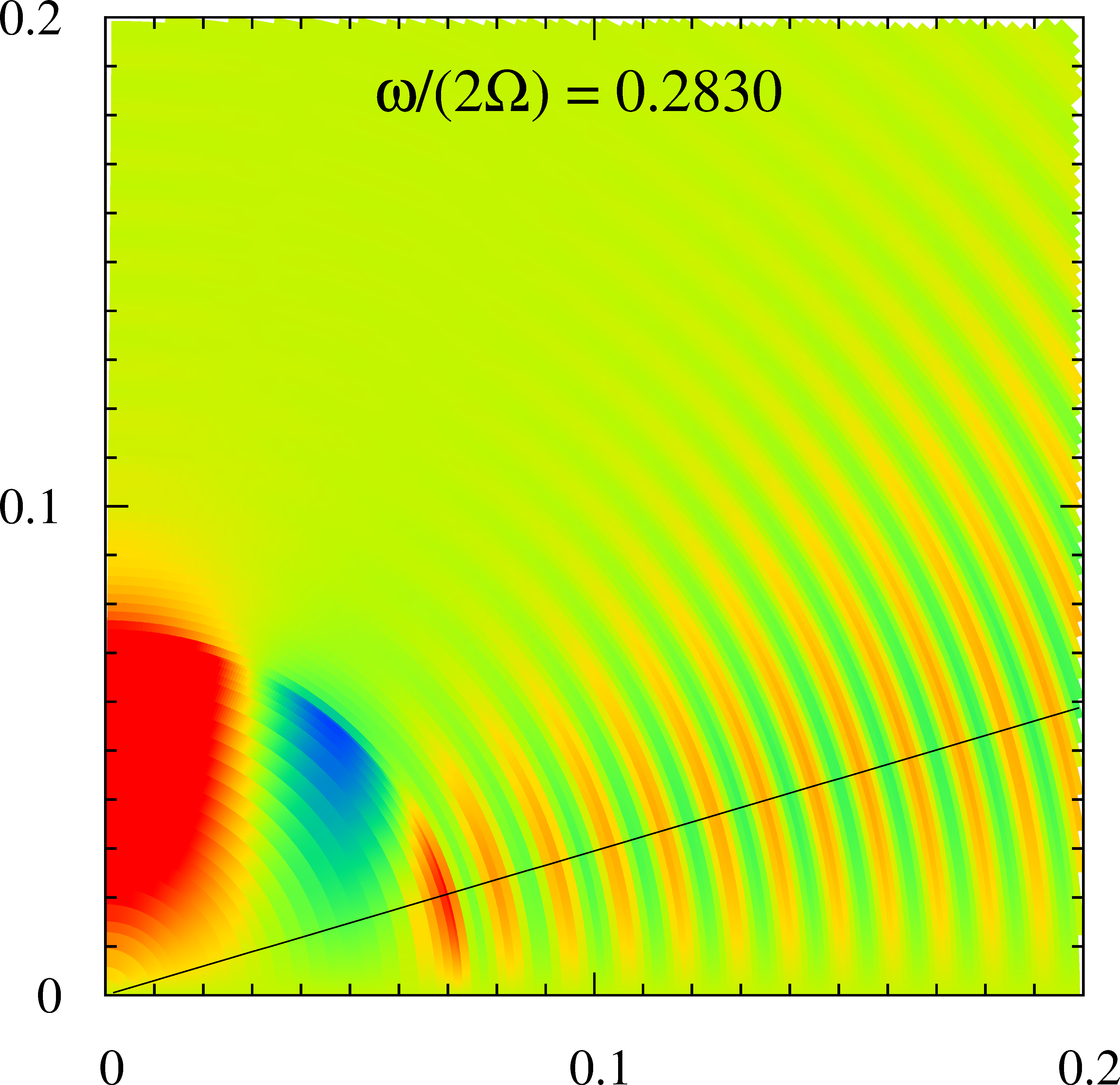}  \hspace*{-0.cm}   \includegraphics[width=0.31\linewidth]{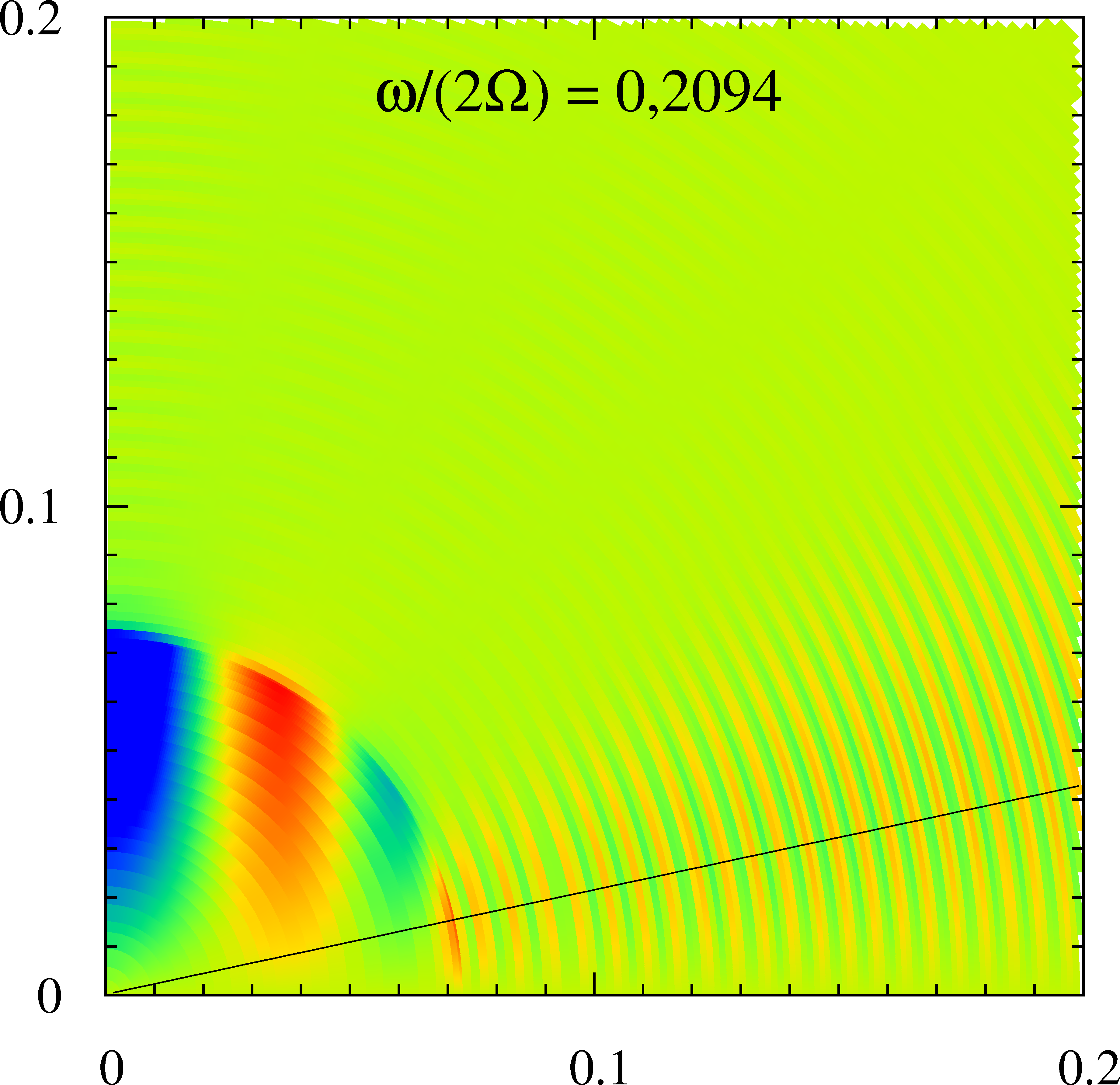}
        \caption{\label{fig:GI_ACOR_2D} Eigenfunction ($P'/\sqrt{\rho}$, perturbed pressure over the equilibrium density) in the meridional plane for the three modes located at resonance in Fig.\ref{fig:DeltaP_m1.40_21muHz}: dipolar zonal modes computed with the {\sc acor} code for model 1z, for a uniform rotation frequency of 21 $\mu$Hz. The left pannel corresponds to the first resonance, for the mode $n=-33$, the middle panel for the mode $n=-73$, and the right panel for the mode $n=-133$. The black solid line corresponds to the critical co-latitude $\theta_c=arccos(\omega_{co}/2\Omega)$, geometrical feature found in pure inertial modes as defined in Sect. \ref{Ss:equations}.}
\end{figure*}
\begin{table}
	\centering
	\caption{Parameters of the models explored in detail in this study. They have been named as follows:  the number 1, 2 3 corresponds to the mass (resp. 1.4, 1.6 and 1.86 M$_{\odot}$), then the letter z stands for ZAMS, m for mid main sequence, and t for TAMS. These models were computed with turbulent diffusion (with Co = 700 cm$^2$ s$^{-1}$) and no overshooting. }
	\label{tab:models_table}
	\begin{tabular}{lccccccccr} 
		\hline
		Model name & 1z & 2m & 3t  \\
		\hline
		\, ~ \, $\rm M/M_{\odot}$ \, ~ \, & \, ~ \, 1.40 \, ~ \, & \, ~ \, 1.60 \, ~ \, & \, 1.86 \\
		\, ~ \, T$_{eff}$ \, ~ \, & 6880 & 7190 & 6760 \\
		\, ~ \, $\log L/L_{\odot}$ \, ~ \, & 0.627 & 1.036 & 1.355 \\
		\, ~ \, $\log$ g \, ~ \, & 4.30 & 3.98 & 3.63 \\
		\, ~ \, $\rm R/R_{\odot}$ \, ~ \, & 1.38 & 2.13 & 3.48 \\
		\, ~ \, Age (Myr) \, ~ \, & 185 & 1830 & 1480 \\
		\, ~ \, X$_{c}$ \, ~ \, & 0.68 & 0.35 & 0.06 \\
		\hline
	\end{tabular}
\end{table}
The oscillation modes were computed as the adiabatic response of the structure to small perturbations,  i.e. of the density, pressure, gravitational potential, and velocity field. The {\sc acor} code was developed for this purpose and is presented in \cite{Ouazzani2012b} and \cite{Ouazzani2015}.
This oscillation code solves the hydrodynamics equations perturbed by Eulerian fluctuations, performing direct integration of the 2D problem. The numerical method is based on a spectral multi-domain method which expands the angular dependence of eigenfunctions onto spherical harmonics series, and whose radial treatment is particularly well adapted to the behaviour of equilibrium quantities in evolved models (at the interface of convective and radiative regions, and at the stellar surface). The radial differentiation scheme is made by means of a sophisticated finite difference method, which is accurate up to the fifth order in terms of the radial resolution \citep[developed by][for the {\sc losc} adiabatic code]{Scuflaire2008a}.

According to \cite{Ballot2012} and \cite{Ouazzani2017}, the 1D non-perturbative approach -- which presents the advantage of requiring less numerical resources --, give satisfactory results compared to the full 2D approach. Throughout this study we therefore work with one-dimensional spherical stellar models. The pulsation modes have been computed using up to 5 spherical harmonics, i.e. for a given $m$,\, $\ell$=1 to 9 with odd value for dipolar modes. We have made sure, by the mean of convergence tests, that the frequencies do not vary significantly when adding a sixth spherical harmonic in the series. The resulting modes have multiple $\ell$ characters, they are then assigned an effective angular degree, taking the dominant contribution to the kinetic energy in the series.

\subsection{Gravito-inertial resonances in theoretical spectra}
\label{Ss:syntheticspectra}

\begin{table}[h!]
  \centering
  \caption{Characteristics of the resonances for zonal dipolar modes for the three models 1z, 2m, 3t given in Tab.\ref{tab:models_table} (radial order $n$, frequency of the mixed gravito-inertial/pure inertial mode $\nu_{in}$, and corresponding spin parameter $s$) and for three uniform rotation frequencies 15, 19 and 23$\mu$Hz. }
  \label{tab:resonances_zonals}
  \begin{tabular}{llllllllllc}
    \hline
    \multicolumn{2}{l}{$\nu_{rot}$\, ~ } & \, ~ 1z \, ~ &\, ~  2m \, ~ & \, ~ 3t \, ~  \\ \hline
    \multirow{3}{*}{15$\mu$Hz\, ~ } & \, ~ $n$ \, ~ & \, ~ -43 \, ~ &\, ~ -40 \, ~ & \, ~ -45\, ~ \\
                              & $\nu_{in}$($\mu$Hz) & \, ~ 13.30 \, ~ & \, ~ 13.89 \, ~ & \, ~ 13.93\, ~ \\
                              & \, ~ $s$ \, ~ & \, ~ 2.256 \, ~ & \, ~ 2.160 \, ~ & \, ~ 2.153\, ~ \\ 
    \hline
    \multirow{3}{*}{19$\mu$Hz\, ~ }& \, ~ $n$ \, ~ & \, ~ -34 \, ~ & \, ~ -32 \, ~ & \, ~ -35 \, ~ \\
                              & $\nu_{in}$($\mu$Hz) & \, ~ 16.81 \, ~ & \, ~ 17.52 \, ~ & \, ~ 17.72 \, ~ \\
                              & \, ~ $s$ \, ~ & \, ~ 2.261 \, ~ & \, ~ 2.169 \, ~ & \, ~ 2.145 \, ~ \\
    \hline
    \multirow{3}{*}{23$\mu$Hz\, ~ }& \, ~ $n$ \, ~ & \, ~ -29 \, ~ & \, ~ -26\, ~ &\, ~ -29 \, ~ \\
                              & $\nu_{in}$($\mu$Hz) & \, ~ 20.04 \, ~ & \, ~ 21.29 \, ~ & \, ~ 21.15 \, ~ \\
                              & \, ~ $s$ \, ~ & \, ~ 2.296 \, ~ & \, ~ 2.161 \, ~ & \, ~ 2.149 \, ~ \\ 
    \hline
    \hline
  \end{tabular}
\end{table}
\begin{table}[h!]
  \centering
  \caption{Same as Tab.\ref{tab:resonances_zonals} for retrograde dipolar modes.}
  \label{tab:resonances_retro}
  \begin{tabular}{llllllllllc}
    \hline
    \multicolumn{2}{l}{$\nu_{rot}$\, ~ } & \, ~ 1z \, ~ &\, ~  2m \, ~ & \, ~ 3t \, ~  \\ \hline
    \multirow{3}{*}{15$\mu$Hz\, ~ } & \, ~ $n$ \, ~ & \, ~ -37 \, ~ &\, ~ -38 \, ~ & \, ~ -42\, ~ \\
                              & $\nu_{in}$($\mu$Hz) & \, ~ 7.35 \, ~ & \, ~ 7.24 \, ~ & \, ~ 7.33\, ~ \\
                              & \, ~ $s$ \, ~ & \, ~ 1.342 \, ~ & \, ~ 1.349 \, ~ & \, ~ 1.344\, ~ \\ 
    \hline
    \multirow{3}{*}{19$\mu$Hz\, ~ }& \, ~ $n$ \, ~ & \, ~ -29 \, ~ & \, ~ -29 \, ~ & \, ~ -33 \, ~ \\
                              & $\nu_{in}$($\mu$Hz) & \, ~ 9.35 \, ~ & \, ~ 9.38 \, ~ & \, ~ 9.27 \, ~ \\
                              & \, ~ $s$ \, ~ & \, ~ 1.340 \, ~ & \, ~ 1.339 \, ~ & \, ~ 1.344 \, ~ \\
    \hline
    \multirow{3}{*}{23$\mu$Hz\, ~ }& \, ~ $n$ \, ~ & \, ~ -25 \, ~ & \, ~ -24\, ~ &\, ~ -28 \, ~ \\
                              & $\nu_{in}$($\mu$Hz) & \, ~ 10.74 \, ~ & \, ~ 11.23 \, ~ & \, ~ 10.49 \, ~ \\
                              & \, ~ $s$ \, ~ & \, ~ 1.363 \, ~ & \, ~ 1.344 \, ~ & \, ~ 1.374 \, ~ \\ 
    \hline
    \hline
  \end{tabular}
\end{table}
\begin{table}[h!]
  \centering
  \caption{Same as Tab.\ref{tab:resonances_zonals} for prograde dipolar modes, and a uniform rotation frequency of 25$\mu$Hz. }
  \label{tab:resonances_pro}
  \begin{tabular}{llllc}
    \hline
    \multicolumn{2}{l}{$\nu_{rot}$\, ~ } & \, ~ 1z \, ~ &\, ~  2m \, ~ & \, ~ 3t \, ~  \\ \hline
    \multirow{3}{*}{25$\mu$Hz\, ~ } & \, ~ $n$ \, ~ & \, ~ -53 \, ~ &\, ~ -44 \, ~ & \, ~ -48\, ~ \\
                              & $\nu_{in}$($\mu$Hz) & \, ~ 29.59 \, ~ & \, ~ 30.65 \, ~ & \, ~ 30.83\, ~ \\
                              & \, ~ $s$ \, ~ & \, ~ 10.891 \, ~ & \, ~ 8.854 \, ~ & \, ~ 8.576\, ~ \\ 
    \hline
    \hline
  \end{tabular}
\end{table}

Using the set up described in Sect. \ref{Ss:models} we have computed synthetic oscillation spectra for dipolar gravito-inertial modes, assuming a set of uniform rotation values ranging from 7$\mu$Hz to 25$\mu$Hz. In Fig. \ref{fig:DeltaP_m1.40_allrot} we illustrate the period spacing of such synthetic spectra for zonal modes with radial orders ranging from -15 to -49. The pure inertial/gravito-inertial resonances appear clearly in the period spacing series as dips, when a few modes show lower period spacing than their neighbouring modes, for the spectra corresponding to rotation frequencies of 13, 15, 17, 19, 21, and 23 $\mu$Hz. The vertical lines in Fig.\ref{fig:DeltaP_m1.40_allrot} stand at the period of the pure inertial modes computed with the method described in Sect. \ref{Ss:equations}. Given the approximations used for their calculations, the agreement with complete calculations is rather stunning, all the more for the lowest rotation rates.
The characteristics of the resonances found in the range of excited modes in the complete synthetic spectra of dipolar modes are given in Tab. \ref{tab:resonances_zonals}, \ref{tab:resonances_retro}, \ref{tab:resonances_pro}, for different rotation rates. The interpretation of the discrepancies between the analytical model presented in Sect.\ref{S:theory}, and the complete ones is reported to Sect.\ref{S:stratif}.

Concerning prograde modes, which are those that are more likely observed, the resonances only appear at high rotation rates. For the three models explored in this study, the value of the uniform rotation rate has to be increased up to 25$\mu$Hz in order to find a resonance among the modes which are likely to be excited (radial orders around -15 to -60). A particular attention should be paid to this case, as the literature has mostly reported observations of prograde modes in $\gamma$ Doradus and SPB stars. In particular, we believe that the phenomenon of gravito-inertial resonance is responsible for the dips in the period spacing series of dipolar prograde modes in KIC 5608334, reported in \cite{Saio2018a}. This can be further confirmed by looking at the value of the spin parameter at which this resonance occurs in KIC 5608334: $s=9.01$, which is compatible with the values presented in Tab.\ref{tab:resonances_pro}.

While relaxing the constraint of mode excitation, we seek for the next resonances given in Tab.\ref{tab:resonances_zonals}, to confirm that they originate from the coupling with pure inertial modes. We therefore have extended the complete calculations to a broader range of radial orders, from -11 to -191. The spin parameters relative to these eigenvalues are given in Fig. \ref{fig:DeltaP_m1.40_21muHz}. As shown in Fig.\ref{fig:DeltaP_m1.40_21muHz}, three resonances are found, and they correspond to the three eigenvalues of the Poincaré problem (indicated in Fig.\ref{fig:DeltaP_m1.40_21muHz} by the grey vertical lines), for which $\zeta_{I}^{GI}$ is not negligible (see Tab.~\ref{table:1}).
In other words, for each eigenvalue with a coupling coefficient of the order of 0.1 or higher, an associated resonance has been found in the complete calculations. 
In Fig.\ref{fig:GI_ACOR_2D} are given the 2D maps of the eigenfunctions of the three modes located in the dips of the period spacing series Fig.\ref{fig:DeltaP_m1.40_21muHz}. These have to be compared to the three panels giving the 2D maps of the pure inertial eigenmodes computed in Sect.\ref{S:theory}, provided in Fig.\ref{fig:GI_HomoSPhere}. From this comparison we see clearly that the modes trapped in the convective cores in Fig.\ref{fig:GI_ACOR_2D} follow closely the morphology of the pure inertial modes in a full sphere of uniform density computed in Sect. \ref{S:theory}.
The mixed morphology of the complete modes clearly confirms their mixed nature: pure inertial in the convective core, and gravito-inertial in the surrounding radiative zone.

\subsection{Failure of the traditional approximation to reproduce the resonances}
\label{Ss:tar}
%
In rotating stars, the equation system for pulsations is not separable in the radial and latitudinal coordinates.
The TAR is an approximate treatment that conserves the separability of the system. The first hypothesis is made on the rotation profile by assuming solid-body rotation. The centrifugal distortion is neglected and hence spherical symmetry is assumed. Furthermore, considering the properties 
of low frequency high order g-modes, the TAR neglects the Coriolis force associated with radial motion and radial component of the Coriolis force associated with horizontal motion. Practically, it consists of neglecting the horizontal component of the angular velocity so that ${\bf\Omega} = \left[\Omega \cos \theta, 0, 0 \right]$ in the spherical polar coordinates. Finally, the Cowling approximation is made \citep{Cowling1941}, which neglects the perturbation of the gravitational potential. As a result, the motion equation for pulsations can be reduced to an equation for the radial component, which is similar to the one without rotation, and a Laplace tidal equation for the horizontal component, whose eigenfunctions are the Hough functions. For a detailed derivation of these equation, we refer to \cite{Unno1989} Sect. 34.3, or \cite{Lee1987a}.

This approximation has been implemented in the {\sc losc} adiabatic and {\sc mad} non-adiabatic oscillation codes, the details of which have been given by \cite{Bouabid2013} and \cite{Salmon2014}. 
We have computed a selection of spectra with this method, for the sake of comparison with the non-perturbative approach, for zonal modes in the model 1z, and for the whole rotation frequency range from 7 to 23 $\mu$Hz. The results are plotted in Fig. \ref{fig:DeltaP_m1.40_allrot}, as the grey solid lines. It appears from that figure that the TAR does not model correctly the modes in the neighbourhood of the resonance. While it is in very good agreement far from resonance, it completely misses the change of nature of the mode which endorses a gravito-inertial/pure inertial mixed character.

This result is not surprising since the TAR only applies when $\Omega\ll N$. Thus, in a convective zone where $N=0$, the use of the TAR is not justified. Moreover, they cannot compute pure inertial modes which are not separable in $r$ and $\theta$. Then, as long as the mode is confined in the radiative zone, the TAR reproduces closely enough the complete calculations, but near the resonances when the pure inertial character is significant in the convective core, the TAR fails as expected. Finally, comparing, as in Fig. \ref{fig:DeltaP_m1.40_allrot}, TAR results with complete calculations can be a way to identify resonances due to pure inertial modes.


\section{Impact of the stellar stratification and the rotation rate on the resonances}
\label{S:stratif}
\begin{figure}[t!]
  \centering
  \includegraphics[width=1\linewidth]{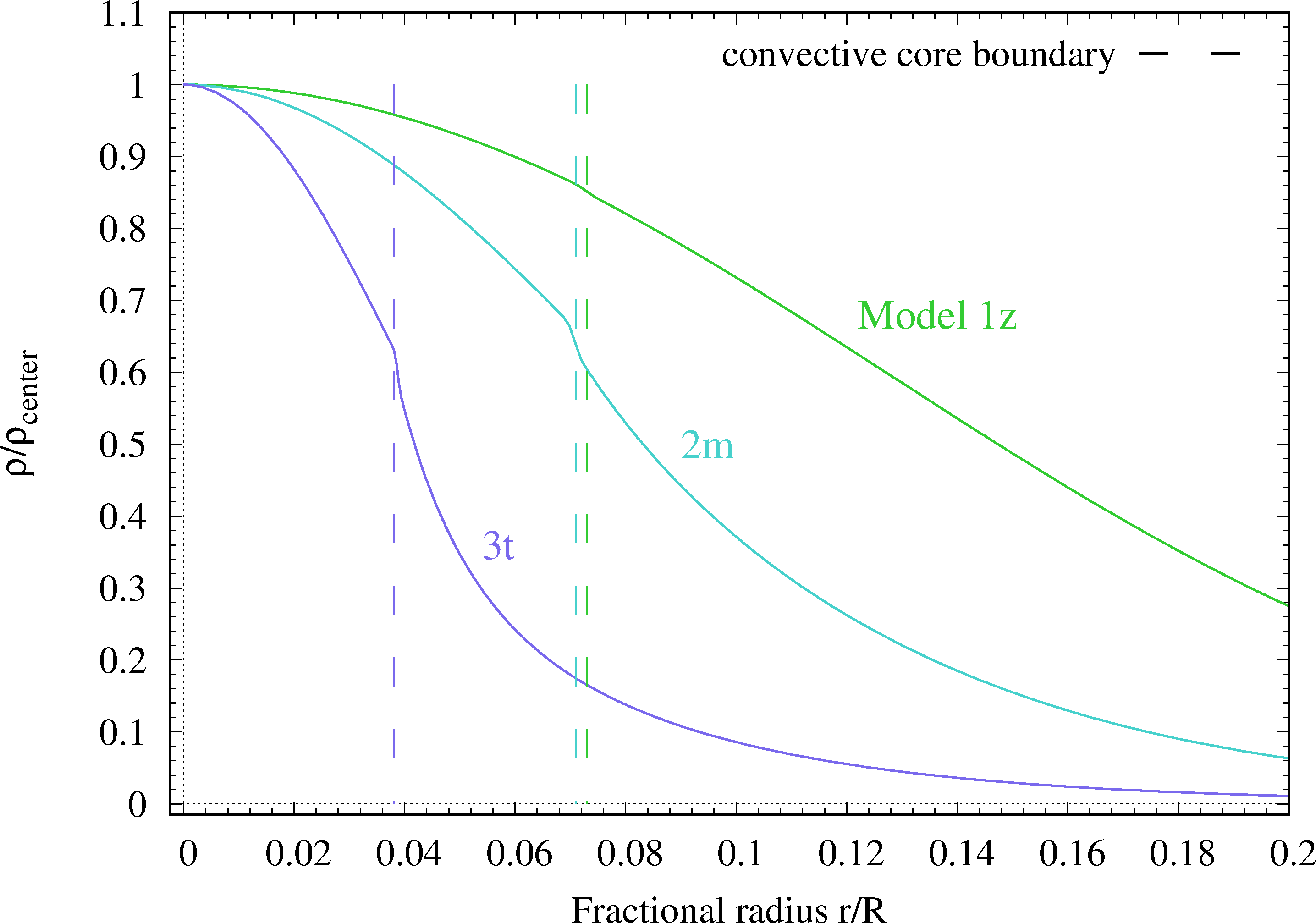}
        \caption{\label{fig:models_density} Density profile, scaled by the central value, for the three stellar models given in Table \ref{tab:models_table}. In dashed line are indicated the external boundary of the convective core for each model.}
 \end{figure}

The density stratification affects inertial waves through the first right hand side term of equation Eq.~(\ref{eq:wave}) which is proportional to the inverse of the density scale height and involves first order derivatives of $\Psi$.
In the short-wavelength limit, second order derivatives dominate over first order ones. Moreover, $H$ does not vanish in stellar cores, thus the density stratification term should have a negligible effect in this limit. However, the short-wavelength limit is not expected to be accurate for the low $\ell_i-|m|$ modes we are interested in, and some effects of the density stratification on the inertial modes are to be expected.

\subsection{Location of the resonances for different stellar models}
We first investigate this effect by exploring the oscillation spectra of stellar models with various density profiles. The three stellar models presented in Tab.\ref{tab:models_table} are investigated. Their stratifications are illustrated through the density profiles plotted in Fig.\ref{fig:models_density}. Indeed the younger the stellar model is, the closer to the homogeneous case. 
In order to find gravito-inertial resonances, we computed synthetic oscillation spectra for dipolar zonal and prograde modes at various uniform rotation rates.

\begin{figure}[t!]
\hspace*{-0.3cm}  \includegraphics[width=1\linewidth]{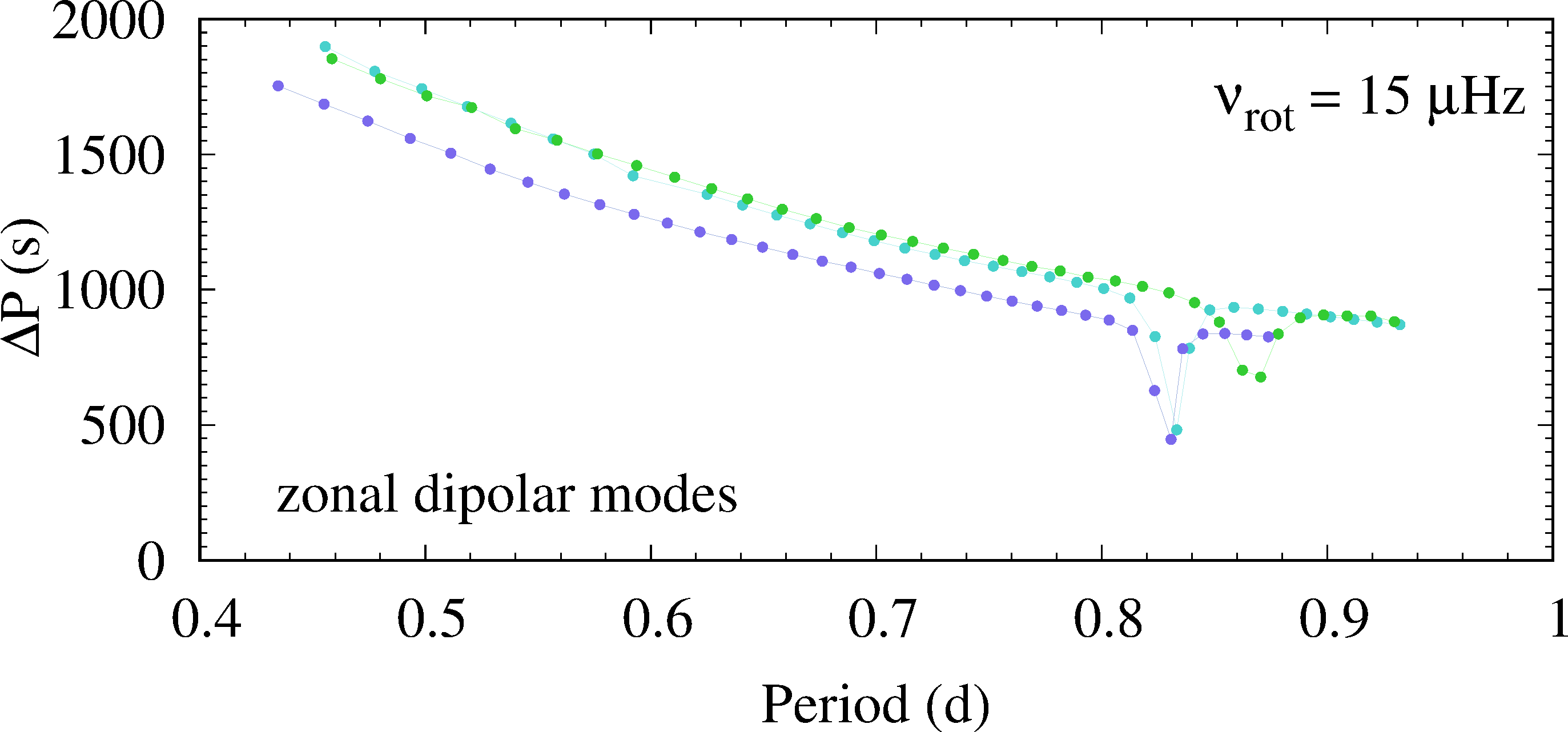}\\
  \includegraphics[width=1\linewidth]{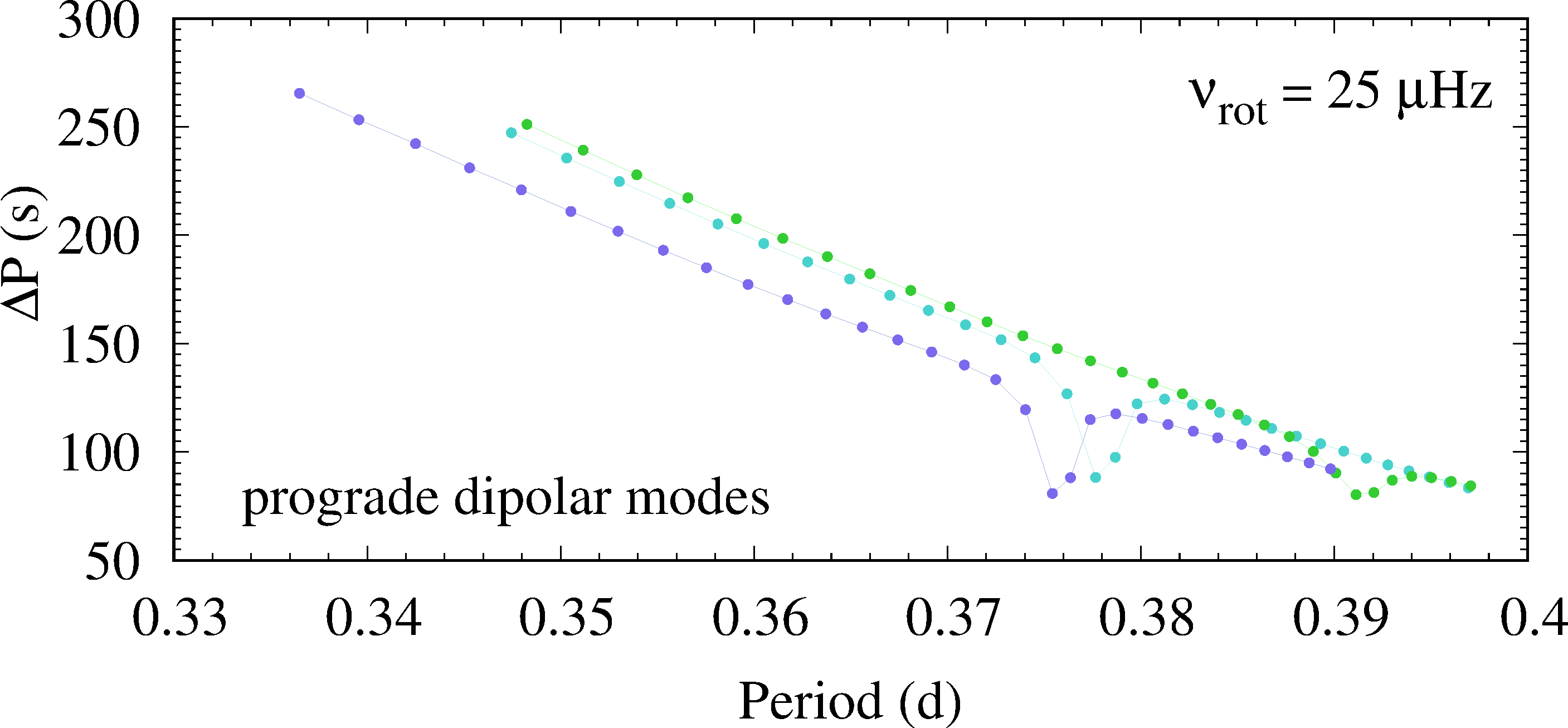}
        \caption{\label{fig:DeltaP_allmod_zonal15} Period spacings as a function of the period for the three stellar models given in Table \ref{tab:models_table}. {\it Top:} for zonal dipolar modes with uniform rotation of 15 $\mu$Hz. Bottom: for prograde dipolar modes with uniform rotation of 25 $\mu$Hz. Both the period spacings and the periods are shown in the inertial frame of reference.}
\end{figure}

For dipolar zonal modes, as well as dipolar retrograde modes, resonances appear at relatively slow rotation, as illustrated Fig.\ref{fig:DeltaP_allmod_zonal15}, and as given in Tab.\ref{tab:resonances_zonals} and Tab.\ref{tab:resonances_retro}. Whereas for prograde modes, as mentioned above, one has to push as high as 25$\mu$Hz to have the resonance occur in the range of frequency where modes are excited. These values have to be compared to the case of pure inertial mode frequencies computed in a homogeneous sphere given in Tab.\ref{table:1}: the strongest resonance, in that case occurs at a spin parameter of  $s=1.3246$ for the retrograde modes, $s=2.2361$ for the zonal modes, and $s=11.3245$ for the prograde ones. For the three classes of modes, it appears clearly, as expected, that the resonances are closer to the homogeneous case for the youngest stellar model, than for the second youngest, than for the oldest, ranging from a difference of less than 1\% to 3\% in the zonal case, 1\% to 2\% in the retrograde case, and 4\% to 24\% in the prograde case.

The discrepancy between the homogeneous case and the stellar model case remains relatively small for the retrograde and zonal modes whereas the resonance of the prograde dipolar mode seems to be more sensitive to the density stratification. As mentioned above, pure inertial modes with wavelengths much shorter than the density scaleheight $H$ should not be strongly affected by the density stratification. The wavelengths of the low degree inertial modes considered here are of the order of a fraction of the convective core radius $r_{cc}$. They are still shorter than the density scaleheight because $r_{cc} \sim H$. This may explain the relatively small effect of the density stratification for retrograde and zonal modes. However, this argument does not hold for prograde inertial modes because their spatial distribution is highly anisotropic with negligible variations in the vertical direction, such behaviour being related to their high spin parameter. Indeed, estimating the effect of the density stratification by comparing left hand side and right hand terms in Eq. 1., we find that the ratio of the second left hand side term to the second right hand side term  is $\sim 1/(k_z H)$ where $k_z$ the vertical component of the wavevector. This ratio and thus the effect of the density stratification can not be neglected when $k_z$ nearly vanishes, which occurs when the spin parameter is high. In our case, this concerns the three prograde inertial modes with significant coupling listed in Tab.\ref{table:1}. 


The behavior of these inertial/gravito-inertial resonances with rotation is explored in more details next section. 

\subsection{Pure inertial modes in a sphere with density stratification}
\label{Ss:convcore}

Discrepancies between the homogeneous case and the real case can arise either from the difference in stratification, or else from the very crude boundary condition which has been applied at the surface of the sphere in the homogeneous case (Sect. \ref{S:theory}).
In order to isolate the relevant physical effects, a modified version of the {\sc acor} code has been used, which allows the computation of pure inertial modes in stellar models truncated above the convective cores. The principle is that the convective core is extracted from a stellar model, and the pulsations are then calculated in the complete non-perturbative approach. We imposed a rigid condition at the surface of the convective core which ensures that the radial displacement goes to zero: $\xi_r(r=r_{cc})=0$. This condition allows the resolution of the core problem without accounting for the radiative envelope.

This modification of the {\sc acor} code allows us to compute pure inertial modes in truncated convective cores taken from the models in Tab.\ref{tab:models_table}. The results are given in Fig.\ref{fig:compareCC-complete}, where the eigenfrequencies computed for the truncated convective cores, converted into spin parameters (in violet), are compared to the values obtained for the resonances in the complete model (in pink), and to the values obtained solving Eq.\ref{eq:wave} (also given in Tab.\ref{table:1}, in grey). The first striking observation, concerning the values of spin parameters obtained in the convective core models, is their constant nature: they do not vary with rotation. We will come back to this in the following subsection. 
Moreover, as expected, the more evolved the model is --hence the more stratified-- the further away the spin parameters for the truncated core models are from the uniform density case.  

From Fig.\ref{fig:compareCC-complete}, it appears clearly that the resonances in the complete models occur systematically much closer to the values corresponding to eigenfrequencies of pure inertial modes determined in the convective core models than in the homogeneous sphere. This does not come as a surprise, but the differences here exposed show that the largest discrepancies between the homogeneous case and the complete one come from the hypothesis of homogeneity rather than from the crude outer boundary condition ($\xi_r =0$ at the surface of the box), which seems to affect the results marginally. 

\subsection{Impact of the rotation rate}
\label{Ss:Rot}

We have noted in the previous section that the spin parameter at which the resonance between the pure inertial modes and the gravito-inertial modes does not vary with rotation. This is a direct consequence of the fact that in isentropic medium waves are governed by Eq.~\ref{eq:wave}. Indeed, for frequencies much smaller than the acoustic frequencies, the last right hand side term of Eq.~\ref{eq:wave} is negligible and the wave equation only depends on the spin parameter. Hence, in  Tab. \ref{tab:resonances_cc} is given one single value of the spin parameter for each models and for each type of modes. 

\begin{figure}
 \hspace*{-0.1cm} \includegraphics[width=1\linewidth]{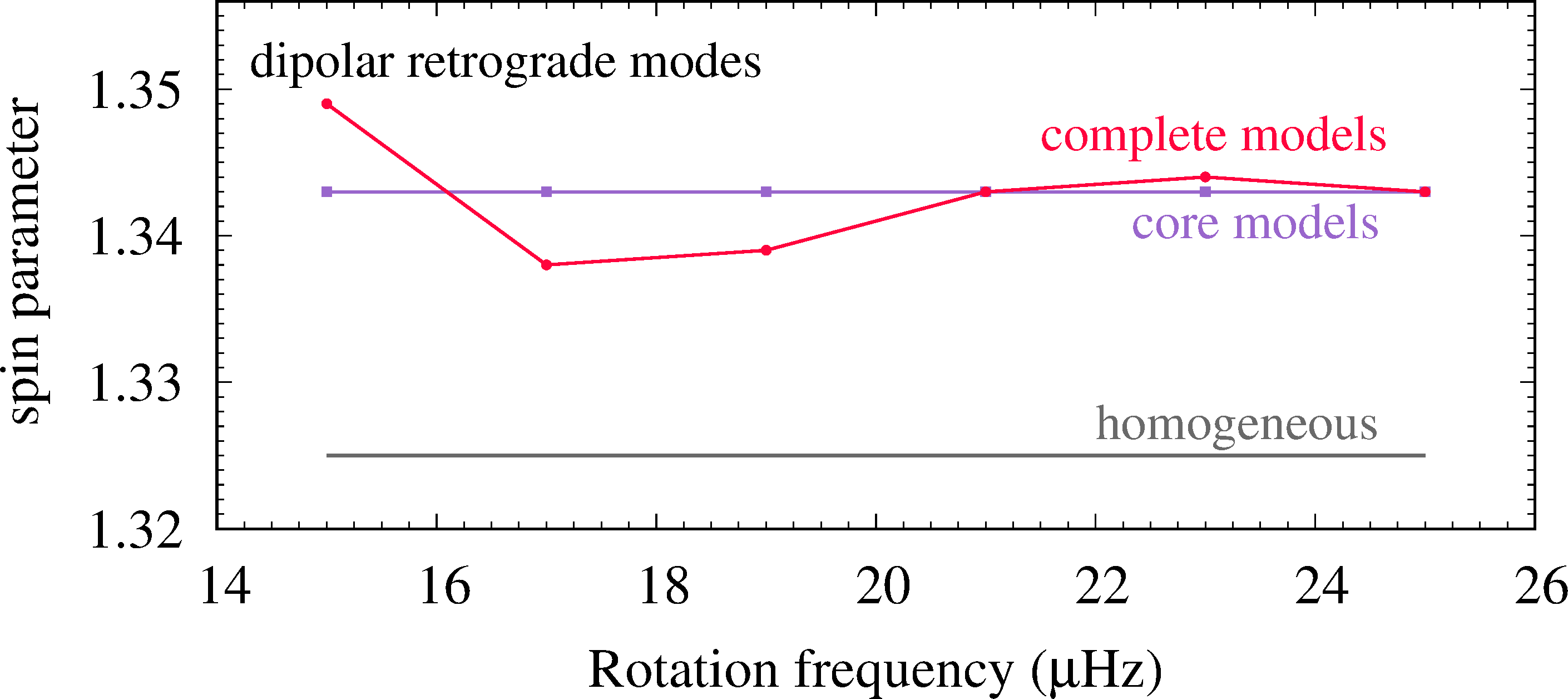}\\
 ~\vspace*{-0.2cm}\\
 \hspace*{-0.1cm} \includegraphics[width=1\linewidth]{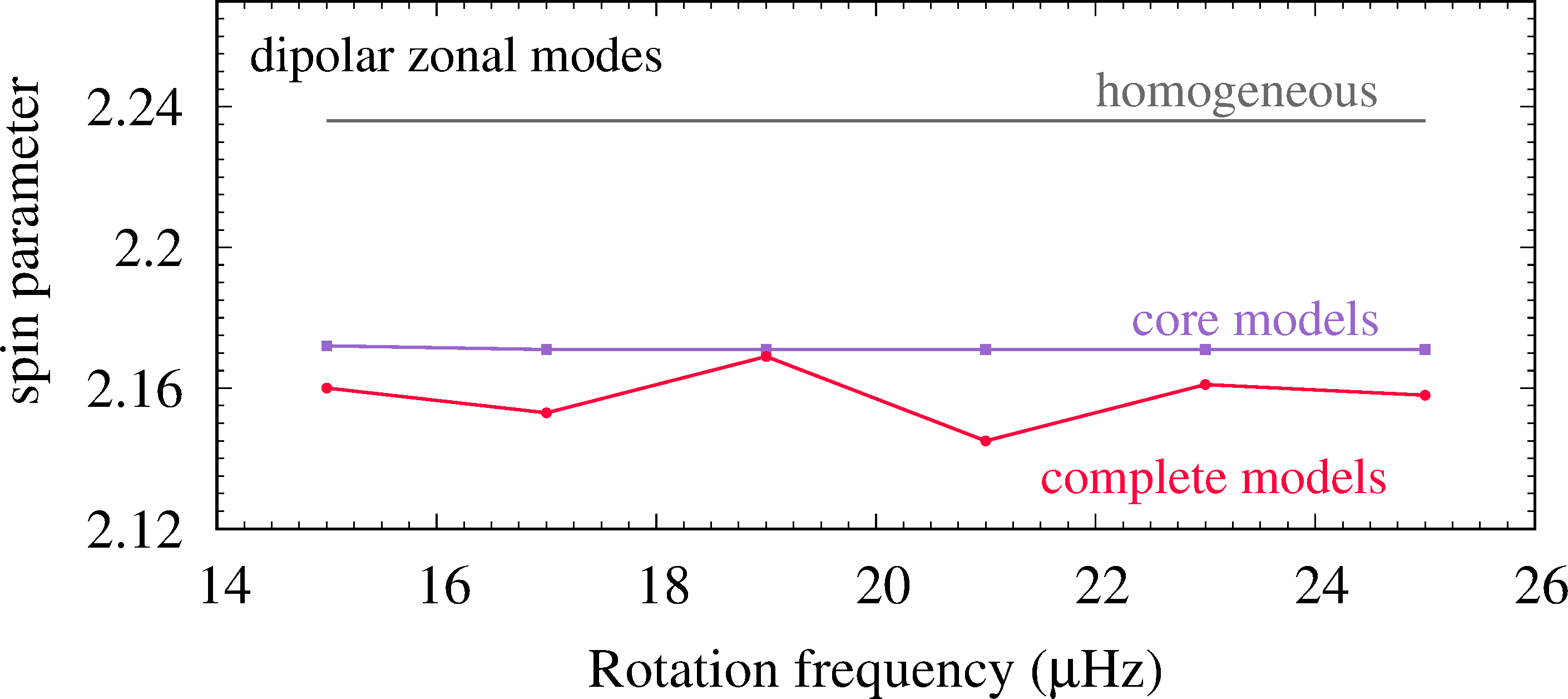}\\
 ~\vspace*{-0.2cm}\\
 \hspace*{0.25cm} \includegraphics[width=0.95\linewidth]{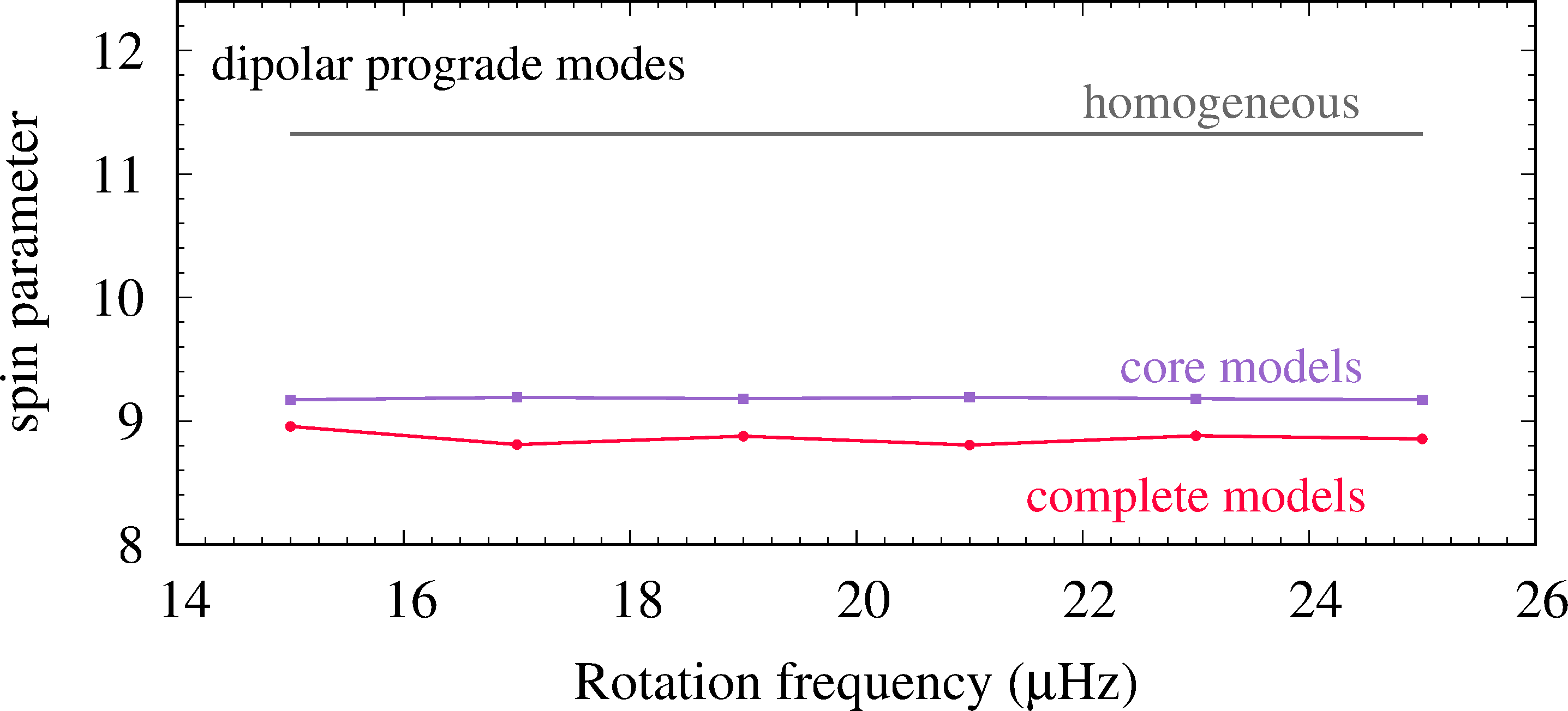}\\
\caption{\label{fig:compareCC-complete} Values of spin parameters as a function of rotation frequency determined through three different methods. In pink are plotted the values computed with {\sc acor} on the stellar model 2m (see Sect. \ref{Ss:syntheticspectra}), the violet curves correspond to values computed in the truncated convective core of model 2m (see Sect. \ref{Ss:convcore}), and the grey constant lines stand for the homogeneous spherical box case (see Sect. \ref{Ss:equations}). These are given for dipolar retrograde modes (upper pannel), zonal (middle), and prograde ones (lower pannel).}
\end{figure}

This is not exactly the case for the complete models. The correct treatment at a boundary separating two distinct propagation cavities is to ensure the continuity of both pressure and radial displacement. \citet{Ogilvie2004} discussed this point at a convective/radiative interface. Arguing that in their respective cavities radial displacements of inertial modes are much larger than those of gravito-inertial modes limited by the Brunt-V\"ais\"al\"a frequency, they deduced from the continuity condition that inertial waves see a $\xi_r \approx 0$ condition at the cavity interface. This justification of the boundary condition is only approximate though. In a more accurate model, the conditions at the interface will also depend on the gravito-inertial mode above, which means that the effective boundary condition for the inertial mode and thus its eigenfrequency will depend on parameters such as the Brunt Vaisala frequency. As a result, the resonance will not occur at a fixed spin parameter but will also vary with, for example, the ratio of the rotation rate to the Brunt-Vaisala frequency. In agreement with this discussion, Fig.~\ref{fig:compareCC-complete} shows variables spin parameters in full calculations. These variations remain relatively small for the stellar models used in our calculations and in the rotation range considered.

\begin{table}
  \centering
  \caption{Eigenfrequencies of pure inertial modes in truncated models of convective cores. These are given for the three models 1z, 2m, 3t given in Tab.\ref{tab:models_table}, and for dipolar retrograde, zonal and prograde modes. }
  \label{tab:resonances_cc}
  \begin{tabular}{llllllllllc}
    \hline
    \, ~ &\, 1z  \, ~ &\, 2m  \, ~ &\, ~ 3t \, ~\\ \hline
    retrograde\, ~ &\, 1.33  \, ~ &\, 1.34  \, ~ &\, ~ 1.35 \, ~\\ \hline
    zonal\, ~ &\, 2.21  \, ~ &\, 2.17  \, ~ &\, ~ 2.16 \, ~\\ \hline
    prograde\, ~ &\, 10.43  \, ~ &\, 9.18  \, ~ &\, ~ 8.83 \, ~\\ \hline
    \hline
  \end{tabular}
\end{table}

 \section{Discussion and conclusions}
 \label{S:cp}


We have studied the resonances which can occur between pure inertial modes trapped in convective cores, and gravito-inertial modes in the surrounding radiative envelopes. We have shown that these resonances leave a clear signature in gravito-inertial modes period spacings which have been found in $\gamma$ Doradus stars observed during the {\it Kepler} nominal mission, including KIC 5608334 \citep{Saio2018a}.

We have proposed a very simple model that provides the spin parameters at which the resonances occur. Comparisons with full calculations show that the model rightly predicts the occurrence of the resonances, while the values of the spin parameters at resonance slightly differ.
One of the assumption of the model, namely the absence of density stratification, has been identified as the main source of discrepancy on the resonance spin parameter. As compared to uniform density model, taking into account density stratification does not change the fact that the resonances occur at fixed spin parameters. 
The other assumption, i.e. vanishing radial displacements at the core radius as a boundary condition for inertial modes, should then account for the remaining deviations with the full calculations. 

In this paper we assumed that the convective core is non-magnetic and uniformly rotating. One could question the relevance of such hypotheses: indeed, convective cores may be differentially rotating, and numerical simulations of A-type star convective cores \citep[see][for B stars]{Brun2005,Augustson2016} show that strong magnetic fields can be generated by a dynamo process. However, in the low Rossby number regime characterizing these convective motions, (i) the maxwell stresses are found to largely suppress the differential rotation \citep[no more than a few percents in][ based on simulations of a 7-days period rotating star]{Brun2005} and (ii) the ratio of the magnetic energy to the kinetic energy of the rotational motions is much smaller than one ($B_{rms}^2/(4 \pi \rho r_c^2 \Omega^2) \sim 2.10^{-4}$ half way through the convective core for the same simulation). The latter property suggests that the magnetic fields generated in the $\gamma$ Dor convective cores have either a very small or a negligible effect on the inertial modes.
Indeed local and global analysis \citep[e.g.][]{Finlay2008,Canet2014} show that the relative deviation of the frequency of an inertial mode provoked by a  magnetic field is proportional to this ratio (when this ratio is small).
Contrary to magnetic fields, even a small differential rotation would matter at least because it will impact the definition of the spin parameter. \cite{Baruteau2013} studied the effect of a differential rotation in the case of an uniform density fluid, and showed that it can strongly modifiy some inertial modes by reducing the volume of their resonant cavity. This effect occurs in specific frequency domains determined by the distribution $\Omega(r,\theta)$. As an additional effect, the non-axisymmetric inertial modes may be affected by the presence of critical layers when their phase speed equals the azimuthal velocity somewhere in their resonant cavity. Assuming a weak cylindrical differential rotation similar to the one found by \cite{Brun2005} for a 7-days rotating A-star, $\Omega/\Omega_0=1 + \eta (r \sin \theta/R)^2$ with $\eta=-0.05$, and applying the criteria given in \cite{Baruteau2013}, we find that the inertial modes producing the dips in the dipolar gravity mode period spacings (see Table \ref{table:1}) are not concerned by these two strong effects of the differential rotation. In order to explore further the influence of differential rotation on the exact spin parameter of the dips, a dedicated study in a stellar context will be necessary.

In future works, it would also be worth investigating whether the dip spin parameters are sensitive to stellar modelling options especially those that modify the structure near the convective core. Furthermore, beyond the central position of the dip, another interesting development concerns its shape and how it depends on the star properties.

In time, one expects to be able to obtain, given one single series of gravito-inertial modes holding the signature of resonance with a pure inertial mode, the rotation rate both in the convective core and in the surrounding radiative near-core region at once. The diagnostic potential of these resonances hold the promise to finally probe the central layers of stars on the main sequence.

In this paper, we focused our attention on mixed pure inertial/dipolar gravito-inertial modes because mostly dipolar-inertial modes have been identified so far. However, resonances will also occur with the other gravito-inertial modes. They can be investigated theoretically as in the present work, using the simplified model and the full calculations. 

\begin{acknowledgements}
This work was supported by the "Programme National de Physique Stellaire" (PNPS) of CNRS/INSU co-funded by CEA and CNES. We thank ISSI (“International Space SciencevInstitute”) through the SoFAR (“Seismology of Fast Rotating Stars”) program for their support. The authors are very grateful to Hideyuki Saio for his great patience. F.L. and J.B. also would like to thank P.-M. Culpin for his significative contribution at early stages of this work during his Master internship.
\end{acknowledgements} 

\bibliographystyle{bibtex/aa}

\begin{thebibliography}{62}
\expandafter\ifx\csname natexlab\endcsname\relax\def\natexlab#1{#1}\fi

\bibitem[{{Angulo} {et~al.}(1999){Angulo}, {Arnould}, {Rayet}, {Descouvemont},
  {Baye}, {Leclercq-Willain}, {Coc}, {Barhoumi}, {Aguer}, {Rolfs}, {Kunz},
  {Hammer}, {Mayer}, {Paradellis}, {Kossionides}, {Chronidou}, {Spyrou},
  {degl'Innocenti}, {Fiorentini}, {Ricci}, {Zavatarelli}, {Providencia},
  {Wolters}, {Soares}, {Grama}, {Rahighi}, {Shotter}, \& {Lamehi
  Rachti}}]{Angulo1999}
{Angulo}, C., {Arnould}, M., {Rayet}, M., {et~al.} 1999, Nuclear Physics A,
  656, 3

\bibitem[{{Asplund} {et~al.}(2009){Asplund}, {Grevesse}, {Sauval}, \&
  {Scott}}]{Asplund2009}
{Asplund}, M., {Grevesse}, N., {Sauval}, A.~J., \& {Scott}, P. 2009, \araa, 47,
  481

\bibitem[{{Augustson} {et~al.}(2016){Augustson}, {Brun}, \&
  {Toomre}}]{Augustson2016}
{Augustson}, K.~C., {Brun}, A.~S., \& {Toomre}, J. 2016, \apj, 829, 92

\bibitem[{{Ballot} {et~al.}(2012){Ballot}, {Ligni{\`e}res}, {Prat}, {Reese}, \&
  {Rieutord}}]{Ballot2012}
{Ballot}, J., {Ligni{\`e}res}, F., {Prat}, V., {Reese}, D.~R., \& {Rieutord},
  M. 2012, in Astronomical Society of the Pacific Conference Series, Vol. 462,
  Progress in Solar/Stellar Physics with Helio- and Asteroseismology, ed.
  H.~{Shibahashi}, M.~{Takata}, \& A.~E. {Lynas-Gray}, 389

\bibitem[{{Baruteau} \& {Rieutord}(2013)}]{Baruteau2013}
{Baruteau}, C. \& {Rieutord}, M. 2013, Journal of Fluid Mechanics, 719, 47

\bibitem[{{Beck} {et~al.}(2012){Beck}, {Montalban}, {Kallinger}, {De Ridder},
  {Aerts}, {Garc{\'{\i}}a}, {Hekker}, {Dupret}, {Mosser}, {Eggenberger},
  {Stello}, {Elsworth}, {Frandsen}, {Carrier}, {Hillen}, {Gruberbauer},
  {Christensen-Dalsgaard}, {Miglio}, {Valentini}, {Bedding}, {Kjeldsen},
  {Girouard}, {Hall}, \& {Ibrahim}}]{Beck2012}
{Beck}, P.~G., {Montalban}, J., {Kallinger}, T., {et~al.} 2012, \nat, 481, 55

\bibitem[{{Belkacem} {et~al.}(2015){Belkacem}, {Marques}, {Goupil}, {Mosser},
  {Sonoi}, {Ouazzani}, {Dupret}, {Mathis}, \& {Grosjean}}]{Belkacem2015b}
{Belkacem}, K., {Marques}, J.~P., {Goupil}, M.~J., {et~al.} 2015, \aap, 579,
  A31

\bibitem[{{B{\"o}hm-Vitense}(1958)}]{Bohm-Vitense1958}
{B{\"o}hm-Vitense}, E. 1958, \zap, 46, 108

\bibitem[{{Bouabid} {et~al.}(2013){Bouabid}, {Dupret}, {Salmon},
  {Montalb{\'a}n}, {Miglio}, \& {Noels}}]{Bouabid2013}
{Bouabid}, M.-P., {Dupret}, M.-A., {Salmon}, S., {et~al.} 2013, \mnras, 429,
  2500

\bibitem[{{Brun} {et~al.}(2005){Brun}, {Browning}, \& {Toomre}}]{Brun2005}
{Brun}, A.~S., {Browning}, M.~K., \& {Toomre}, J. 2005, \apj, 629, 461

\bibitem[{{Bryan}(1889)}]{Bryan1889}
{Bryan}, G.~H. 1889, Philosophical Transactions of the Royal Society of London
  Series A, 180, 187

\bibitem[{{Canet} {et~al.}(2014){Canet}, {Finlay}, \& {Fournier}}]{Canet2014}
{Canet}, E., {Finlay}, C.~C., \& {Fournier}, A. 2014, Physics of the Earth and
  Planetary Interiors, 229, 1

\bibitem[{{Cantiello} {et~al.}(2014){Cantiello}, {Mankovich}, {Bildsten},
  {Christensen-Dalsgaard}, \& {Paxton}}]{Cantiello2014}
{Cantiello}, M., {Mankovich}, C., {Bildsten}, L., {Christensen-Dalsgaard}, J.,
  \& {Paxton}, B. 2014, \apj, 788, 93

\bibitem[{{Castelli} \& {Kurucz}(2003)}]{Castelli2003}
{Castelli}, F. \& {Kurucz}, R.~L. 2003, in IAU Symposium, Vol. 210, Modelling
  of Stellar Atmospheres, ed. N.~{Piskunov}, W.~W. {Weiss}, \& D.~F. {Gray},
  A20

\bibitem[{{Christophe} {et~al.}(2018){Christophe}, {Ballot}, {Ouazzani},
  {Antoci}, \& {Salmon}}]{Christophe2018}
{Christophe}, S., {Ballot}, J., {Ouazzani}, R.-M., {Antoci}, V., \& {Salmon},
  S.~J.~A.~J. 2018, \aap, 618, A47

\bibitem[{{Cowling}(1941)}]{Cowling1941}
{Cowling}, T.~G. 1941, \mnras, 101, 367

\bibitem[{{Deheuvels} {et~al.}(2012){Deheuvels}, {Garc{\'{\i}}a}, {Chaplin},
  {Basu}, {Antia}, {Appourchaux}, {Benomar}, {Davies}, {Elsworth}, {Gizon},
  {Goupil}, {Reese}, {Regulo}, {Schou}, {Stahn}, {Casagrande},
  {Christensen-Dalsgaard}, {Fischer}, {Hekker}, {Kjeldsen}, {Mathur}, {Mosser},
  {Pinsonneault}, {Valenti}, {Christiansen}, {Kinemuchi}, \&
  {Mullally}}]{Deheuvels2012}
{Deheuvels}, S., {Garc{\'{\i}}a}, R.~A., {Chaplin}, W.~J., {et~al.} 2012, \apj,
  756, 19

\bibitem[{{Dintrans} \& {Ouyed}(2001)}]{Dintrans2001}
{Dintrans}, B. \& {Ouyed}, R. 2001, \aap, 375, L47

\bibitem[{{Dintrans} \& {Rieutord}(2000)}]{Dintrans2000}
{Dintrans}, B. \& {Rieutord}, M. 2000, \aap, 354, 86

\bibitem[{{Dziembowski} {et~al.}(1987){Dziembowski}, {Kosovichev}, \&
  {Kozlowski}}]{Dziembowski1987b}
{Dziembowski}, W., {Kosovichev}, A., \& {Kozlowski}, M. 1987, \actaa, 37, 331

\bibitem[{{Eggenberger} {et~al.}(2012){Eggenberger}, {Montalb{\'a}n}, \&
  {Miglio}}]{Eggenberger2012}
{Eggenberger}, P., {Montalb{\'a}n}, J., \& {Miglio}, A. 2012, \aap, 544, L4

\bibitem[{{Ferguson} {et~al.}(2005){Ferguson}, {Alexander}, {Allard}, {Barman},
  {Bodnarik}, {Hauschildt}, {Heffner-Wong}, \& {Tamanai}}]{Ferguson2005}
{Ferguson}, J.~W., {Alexander}, D.~R., {Allard}, F., {et~al.} 2005, \apj, 623,
  585

\bibitem[{{Finlay}(2008)}]{Finlay2008}
{Finlay}, C.~C. 2008, Physics of the Earth and Planetary Interiors, 170, 1

\bibitem[{{Formicola} {et~al.}(2004){Formicola}, {Imbriani}, {Costantini},
  {Angulo}, {Bemmerer}, {Bonetti}, {Broggini}, {Corvisiero}, {Cruz},
  {Descouvemont}, {F{\"u}l{\"o}p}, {Gervino}, {Guglielmetti}, {Gustavino},
  {Gy{\"u}rky}, {Jesus}, {Junker}, {Lemut}, {Menegazzo}, {Prati}, {Roca},
  {Rolfs}, {Romano}, {Rossi Alvarez}, {Sch{\"u}mann}, {Somorjai}, {Straniero},
  {Strieder}, {Terrasi}, {Trautvetter}, {Vomiero}, \&
  {Zavatarelli}}]{Formicola2004}
{Formicola}, A., {Imbriani}, G., {Costantini}, H., {et~al.} 2004, Physics
  Letters B, 591, 61

\bibitem[{{Fuller} {et~al.}(2014){Fuller}, {Lecoanet}, {Cantiello}, \&
  {Brown}}]{Fuller2014}
{Fuller}, J., {Lecoanet}, D., {Cantiello}, M., \& {Brown}, B. 2014, \apj, 796,
  17

\bibitem[{{Fuller} {et~al.}(2019){Fuller}, {Piro}, \& {Jermyn}}]{Fuller2019}
{Fuller}, J., {Piro}, A.~L., \& {Jermyn}, A.~S. 2019, \mnras, 485, 3661

\bibitem[{{Greenspan}(1968)}]{Greenspan1968}
{Greenspan}, H. 1968, {The Theory of Rotating Fuilds}, ed. {Cambridge:
  Cambridge University Press}

\bibitem[{{Guenther} \& {Gilman}(1985)}]{Guenther1985}
{Guenther}, D.~B. \& {Gilman}, P.~A. 1985, \apj, 295, 195

\bibitem[{{Iglesias} \& {Rogers}(1996)}]{Iglesias1996}
{Iglesias}, C.~A. \& {Rogers}, F.~J. 1996, \apj, 464, 943

\bibitem[{{Kurtz} {et~al.}(2014){Kurtz}, {Saio}, {Takata}, {Shibahashi},
  {Murphy}, \& {Sekii}}]{Kurtz2014}
{Kurtz}, D.~W., {Saio}, H., {Takata}, M., {et~al.} 2014, \mnras, 444, 102

\bibitem[{{Lee} \& {Saio}(1987)}]{Lee1987a}
{Lee}, U. \& {Saio}, H. 1987, \mnras, 224, 513

\bibitem[{{Lee} {et~al.}(1992){Lee}, {Strohmayer}, \& {van Horn}}]{Lee1992}
{Lee}, U., {Strohmayer}, T.~E., \& {van Horn}, H.~M. 1992, \apj, 397, 674

\bibitem[{{Li} {et~al.}(2019){Li}, {Van Reeth}, {Bedding}, {Murphy}, \&
  {Antoci}}]{Li2019}
{Li}, G., {Van Reeth}, T., {Bedding}, T.~R., {Murphy}, S.~J., \& {Antoci}, V.
  2019, \mnras, 487, 782

\bibitem[{{Lockitch} \& {Friedman}(1999)}]{Lockitch1999}
{Lockitch}, K.~H. \& {Friedman}, J.~L. 1999, \apj, 521, 764

\bibitem[{{Marques} {et~al.}(2013){Marques}, {Goupil}, {Lebreton}, {Talon},
  {Palacios}, {Belkacem}, {Ouazzani}, {Mosser}, {Moya}, {Morel}, {Pichon},
  {Mathis}, {Zahn}, {Turck-Chi{\`e}ze}, \& {Nghiem}}]{Marques2013}
{Marques}, J.~P., {Goupil}, M.~J., {Lebreton}, Y., {et~al.} 2013, \aap, 549,
  A74

\bibitem[{{Mathis} {et~al.}(2018){Mathis}, {Prat}, {Amard}, {Charbonnel},
  {Palacios}, {Lagarde}, \& {Eggenberger}}]{Mathis2018}
{Mathis}, S., {Prat}, V., {Amard}, L., {et~al.} 2018, \aap, 620, A22

\bibitem[{{Miglio} {et~al.}(2008){Miglio}, {Montalb{\'a}n}, {Noels}, \&
  {Eggenberger}}]{Miglio2008}
{Miglio}, A., {Montalb{\'a}n}, J., {Noels}, A., \& {Eggenberger}, P. 2008,
  \mnras, 386, 1487

\bibitem[{{Mosser} {et~al.}(2012){Mosser}, {Goupil}, {Belkacem}, {Michel},
  {Stello}, {Marques}, {Elsworth}, {Barban}, {Beck}, {Bedding}, {De Ridder},
  {Garc{\'{\i}}a}, {Hekker}, {Kallinger}, {Samadi}, {Stumpe}, {Barclay}, \&
  {Burke}}]{Mosser2012a}
{Mosser}, B., {Goupil}, M.~J., {Belkacem}, K., {et~al.} 2012, \aap, 540, A143

\bibitem[{{Murphy} {et~al.}(2016){Murphy}, {Fossati}, {Bedding}, {Saio},
  {Kurtz}, {Grassitelli}, \& {Wang}}]{Murphy2016}
{Murphy}, S.~J., {Fossati}, L., {Bedding}, T.~R., {et~al.} 2016, \mnras, 459,
  1201

\bibitem[{{Ogilvie} \& {Lin}(2004)}]{Ogilvie2004}
{Ogilvie}, G.~I. \& {Lin}, D.~N.~C. 2004, \apj, 610, 477

\bibitem[{{Ouazzani} {et~al.}(2012){Ouazzani}, {Dupret}, \&
  {Reese}}]{Ouazzani2012b}
{Ouazzani}, R.-M., {Dupret}, M.-A., \& {Reese}, D.~R. 2012, \aap, 547, A75

\bibitem[{{Ouazzani} {et~al.}(2019){Ouazzani}, {Marques}, {Goupil},
  {Christophe}, {Antoci}, {Salmon}, \& {Ballot}}]{Ouazzani2019}
{Ouazzani}, R.~M., {Marques}, J.~P., {Goupil}, M.~J., {et~al.} 2019, \aap, 626,
  A121

\bibitem[{{Ouazzani} {et~al.}(2015){Ouazzani}, {Roxburgh}, \&
  {Dupret}}]{Ouazzani2015}
{Ouazzani}, R.-M., {Roxburgh}, I.~W., \& {Dupret}, M.-A. 2015, \aap, 579, A116

\bibitem[{{Ouazzani} {et~al.}(2017){Ouazzani}, {Salmon}, {Antoci}, {Bedding},
  {Murphy}, \& {Roxburgh}}]{Ouazzani2017}
{Ouazzani}, R.-M., {Salmon}, S.~J.~A.~J., {Antoci}, V., {et~al.} 2017, \mnras,
  465, 2294

\bibitem[{{Papaloizou} \& {Pringle}(1981)}]{Papaloizou1981}
{Papaloizou}, J. \& {Pringle}, J.~E. 1981, \mnras, 195, 743

\bibitem[{{Papaloizou} \& {Savonije}(1997)}]{Papaloizou1997}
{Papaloizou}, J.~C.~B. \& {Savonije}, G.~J. 1997, \mnras, 291, 651

\bibitem[{{Pin{\c{c}}on} {et~al.}(2017){Pin{\c{c}}on}, {Belkacem}, {Goupil}, \&
  {Marques}}]{Pincon2017}
{Pin{\c{c}}on}, C., {Belkacem}, K., {Goupil}, M.~J., \& {Marques}, J.~P. 2017,
  \aap, 605, A31

\bibitem[{{Prat} {et~al.}(2016){Prat}, {Ligni{\`e}res}, \& {Ballot}}]{Prat2016}
{Prat}, V., {Ligni{\`e}res}, F., \& {Ballot}, J. 2016, \aap, 587, A110

\bibitem[{{Rieutord}(1991)}]{Rieutord1991}
{Rieutord}, M. 1991, Geophysical and Astrophysical Fluid Dynamics, 59, 185

\bibitem[{{Rieutord} \& {Valdettaro}(2018)}]{Rieutord2018}
{Rieutord}, M. \& {Valdettaro}, L. 2018, Journal of Fluid Mechanics, 844, 597

\bibitem[{{Rogers} \& {Nayfonov}(2002)}]{Rogers2002}
{Rogers}, F.~J. \& {Nayfonov}, A. 2002, \apj, 576, 1064

\bibitem[{{Saio} {et~al.}(2018){Saio}, {Bedding}, {Kurtz}, {Murphy}, {Antoci},
  {Shibahashi}, {Li}, \& {Takata}}]{Saio2018a}
{Saio}, H., {Bedding}, T.~R., {Kurtz}, D.~W., {et~al.} 2018, \mnras, 477, 2183

\bibitem[{{Saio} {et~al.}(2015){Saio}, {Kurtz}, {Takata}, {Shibahashi},
  {Murphy}, {Sekii}, \& {Bedding}}]{Saio2015}
{Saio}, H., {Kurtz}, D.~W., {Takata}, M., {et~al.} 2015, \mnras, 447, 3264

\bibitem[{{Salmon} {et~al.}(2014){Salmon}, {Montalb{\'a}n}, {Reese}, {Dupret},
  \& {Eggenberger}}]{Salmon2014}
{Salmon}, S.~J.~A.~J., {Montalb{\'a}n}, J., {Reese}, D.~R., {Dupret}, M.-A., \&
  {Eggenberger}, P. 2014, \aap, 569, A18

\bibitem[{{Scuflaire} {et~al.}(2008{\natexlab{a}}){Scuflaire}, {Montalb{\'a}n},
  {Th{\'e}ado}, {Bourge}, {Miglio}, {Godart}, {Thoul}, \&
  {Noels}}]{Scuflaire2008a}
{Scuflaire}, R., {Montalb{\'a}n}, J., {Th{\'e}ado}, S., {et~al.}
  2008{\natexlab{a}}, \apss, 316, 149

\bibitem[{{Scuflaire} {et~al.}(2008{\natexlab{b}}){Scuflaire}, {Th{\'e}ado},
  {Montalb{\'a}n}, {Miglio}, {Bourge}, {Godart}, {Thoul}, \&
  {Noels}}]{Scuflaire2008b}
{Scuflaire}, R., {Th{\'e}ado}, S., {Montalb{\'a}n}, J., {et~al.}
  2008{\natexlab{b}}, \apss, 316, 83

\bibitem[{{Townsend}(2003)}]{Townsend2003}
{Townsend}, R.~H.~D. 2003, \mnras, 340, 1020

\bibitem[{{Unno} {et~al.}(1989){Unno}, {Osaki}, {Ando}, {Saio}, \&
  {Shibahashi}}]{Unno1989}
{Unno}, W., {Osaki}, Y., {Ando}, H., {Saio}, H., \& {Shibahashi}, H. 1989,
  {Nonradial oscillations of stars}

\bibitem[{{Van Reeth} {et~al.}(2018){Van Reeth}, {Mombarg}, {Mathis},
  {Tkachenko}, {Fuller}, {Bowman}, {Buysschaert}, {Johnston}, {Garc{\'\i}a
  Hern{\'a}ndez}, \& {Goldstein}}]{VanReeth2018}
{Van Reeth}, T., {Mombarg}, J.~S.~G., {Mathis}, S., {et~al.} 2018, \aap, 618,
  A24

\bibitem[{{Van Reeth} {et~al.}(2016){Van Reeth}, {Tkachenko}, \&
  {Aerts}}]{VanReeth2016}
{Van Reeth}, T., {Tkachenko}, A., \& {Aerts}, C. 2016, aap, 593, A120

\bibitem[{{Wu}(2005)}]{Wu2005}
{Wu}, Y. 2005, \apj, 635, 674

\bibitem[{{Zahn}(1992)}]{Zahn1992}
{Zahn}, J. 1992, \aap, 265, 115

\end{thebibliography}

\appendix
\newpage
\section{Ellipsoidal coordinates}
\label{app:coord}
The Poincaré equation becomes separable using the ellipsoidal coordinates $(x_1,x_2,\phi)$ where $\phi$ is the azimuthal angle of the spherical coordinates and $x_1$ and $x_2$ relations with the cartesian coordinates read :
\begin{eqnarray}
    x &=& \left(\frac{(1-x_1^2)(1-x_2^2)}{1-\mu^2}\right)^{1/2}  \cos(\phi) \\
    y & =& \left(\frac{(1-x_1^2)(1-x_2^2)}{1-\mu^2}\right)^{1/2}\sin(\phi)\\
    z &=& \frac{x_1 x_2}{\mu}
\end{eqnarray}
\noindent where $x_1 \in [\mu, 1]$ and $x_2 \in [-\mu, \mu]$, and distances have been normalized with the sphere radius.
In these coordinates, the surface of the sphere is given by  $\{x_1 \in [\mu, 1], x_2 = \mu\} \cup \{x_1 = \mu, x_2 \in [-\mu, \mu]\}$.

\section{Inertial modes of the full sphere}
\label{app:modes}

\begin{figure*}
  \centering
  \includegraphics[width=0.22\linewidth]{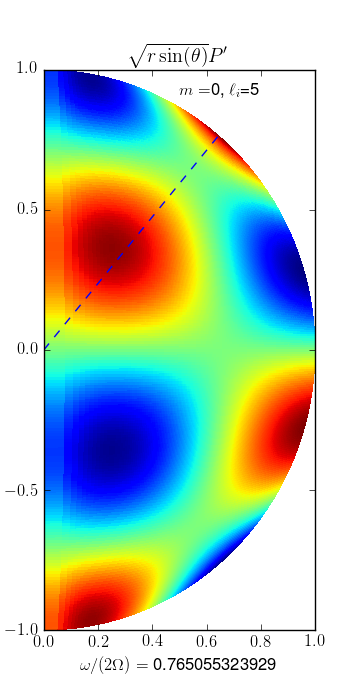}\hspace*{0.7cm}
  \includegraphics[width=0.22\linewidth]{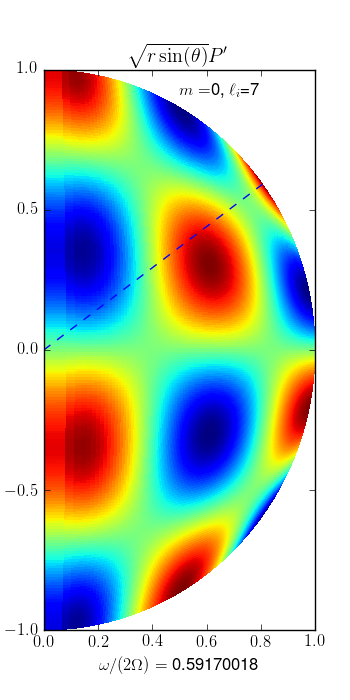}\hspace*{0.7cm}
  \includegraphics[width=0.22\linewidth]{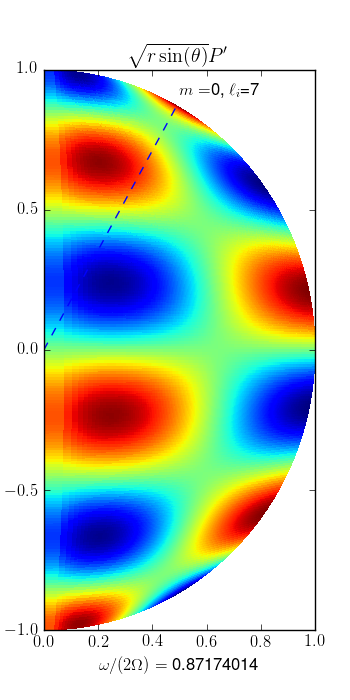}
  \includegraphics[width=0.22\linewidth]{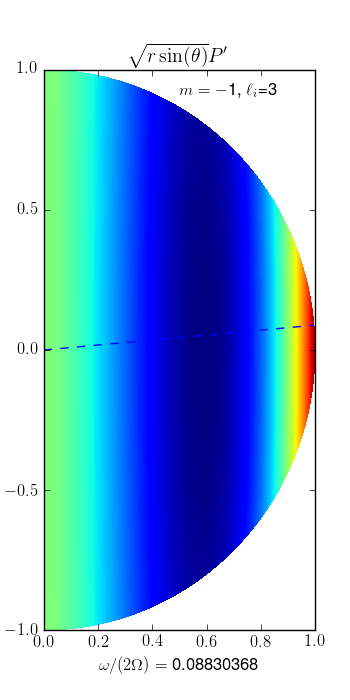}\hspace*{0.7cm}
  \includegraphics[width=0.22\linewidth]{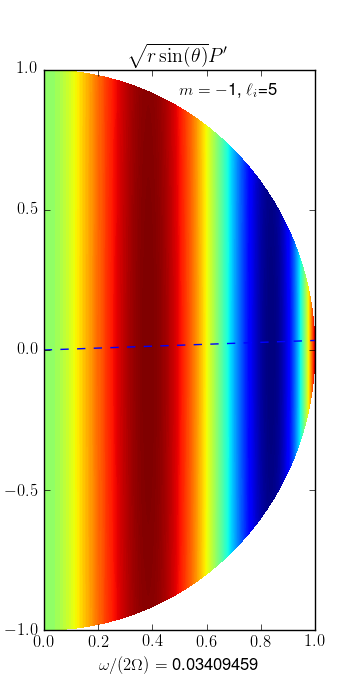}\hspace*{0.7cm}
  \includegraphics[width=0.22\linewidth]{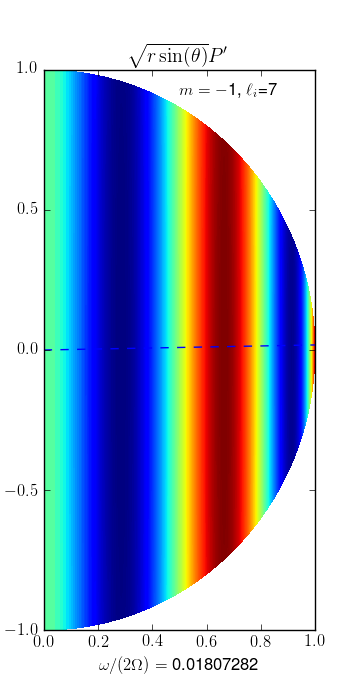}
  \includegraphics[width=0.22\linewidth]{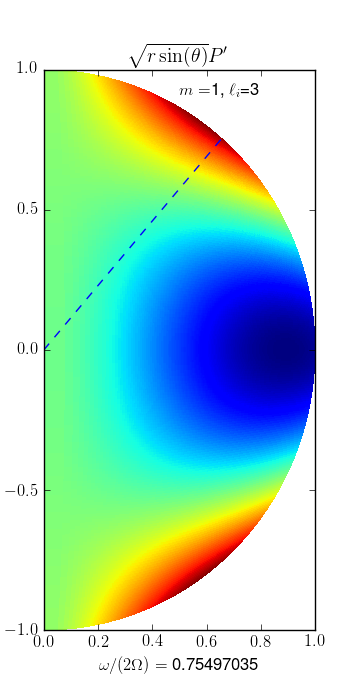}\hspace*{0.7cm}
  \includegraphics[width=0.22\linewidth]{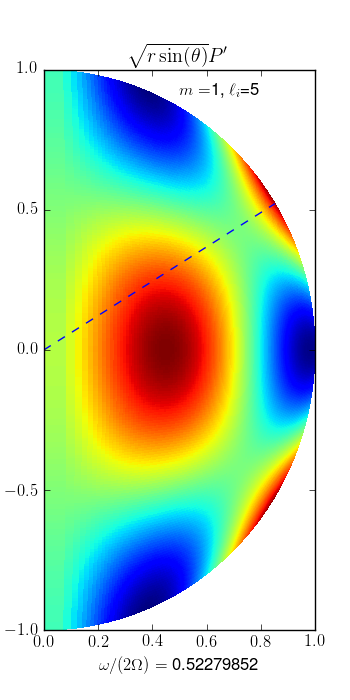}\hspace*{0.7cm}
  \includegraphics[width=0.22\linewidth]{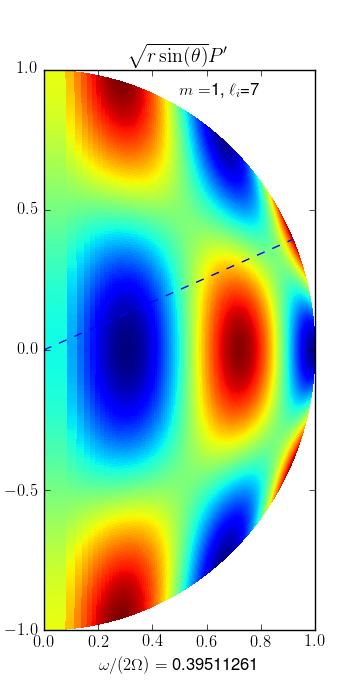}
        \caption{\label{fig:GI_HomoSPhere_App} Inertial modes in a uniform density sphere, as computed with the method given in Sect. \ref{S:theory}. Each of these modes have a small spatial matching with the dipolar gravito-inertial mode of the same spin parameter. The dotted line specifies the critical latitude $\theta_c = \arccos(\omega/(2\Omega))$. The top panels show the zonal modes with low coupling coefficients presented in Tab.\ref{table:1}, whereas the middle and the bottom panels show the tree modes with the highest coupling coefficients for the prograde (m=-1), and the retrograde (m=1) modes. }
\end{figure*}

\end{document}